\documentclass[numberedappendix]{emulateapj}
\newcommand{\Mb}{M_\mathrm{B}}
\newcommand{\mb}{m_\mathrm{B}}
\slugcomment{Accepted for publication by The Astronomical Journal}
\shorttitle{Virgo early-type dwarfs. I. Disk features}
\shortauthors{Lisker, Grebel, \& Binggeli}
\begin{document}
 
\title{Virgo cluster early-type dwarf galaxies with the Sloan Digital
  Sky Survey.\\
I. On the possible disk nature of bright early-type dwarfs
}

\author{Thorsten Lisker, Eva K. Grebel, and Bruno Binggeli}
\affil{Astronomical Institute, Dept.\ of Physics and Astronomy,
  University of Basel, Venusstrasse 7, CH-4102 Binningen, Switzerland}
\email{tlisker@astro.unibas.ch}
 
\begin{abstract}
    We present a systematic search for disk features in 476 Virgo
    cluster early-type dwarf (dE) galaxies. This is the first such
    study of an almost-complete, statistically significant dE sample
    which includes all certain or possible cluster members with $m_{\rm
    B}\le 18$ that are covered by the optical imaging data of the
    Sloan Digital Sky Survey Data Release 4.
    Disk features (spiral arms, edge-on disks, or bars) were identified
    by applying unsharp masks to a combined image from three bands (g,
    r, i), as well as by subtracting the axisymmetric light
    distribution of each galaxy from that image. 14 objects are
    unambiguous identifications of disks, 10 objects show
    'probable disk' features, and 17 objects show 'possible disk'
    features. The number fraction of these galaxies, for which we
    introduce the term dEdi, reaches more than 50\% at the 
    bright end of the dE population, and decreases to less than 5\% for
    magnitudes $m_{\rm B}>16$. Although part of this observed decline
    might be due to the lower signal-to-noise ratio at fainter
    magnitudes, we show that it cannot be caused solely by the
    limitations of our detection method. The luminosity function of
    our full dE sample can be explained by a superposition of dEdis and
    ordinary dEs, strongly suggesting that dEdis are a distinct
    type of galaxy. This is supported by the projected spatial
    distribution: dEdis show basically no clustering and roughly follow
    the spatial distribution of spirals and irregulars, whereas ordinary
    dEs are distributed similarly to the strongly clustered E/S0
    galaxies. While the flattening distribution of ordinary
    dEs is typical for spheroidal objects, the distribution of dEdis is
    significantly different and agrees with their being flat oblate
    objects. We therefore conclude that the dEdis are not spheroidal
    galaxies that just have an embedded disk component, but are instead
    a population of genuine disk galaxies.
    Several dEdis display well-defined spiral arms with grand design
    features that clearly differ from the flocculent, open arms typical
    for late-type spirals that have frequently been proposed as
    progenitors of early-type dwarfs. This raises the question of what
    process is able to create such spiral arms -- with pitch angles
    like those of Sab/Sb galaxies -- in bulgeless dwarf galaxies.
\end{abstract}
 
 

\section{Introduction}
 \label{sec:int}
At first glance, early-type dwarf galaxies (dEs) are characterized by their
smooth appearance, having no recent or ongoing star formation
and apparently no gas or dust content. Since they are the most numerous
type of galaxy in clusters, it is self-evident that most of the
proposed formation scenarios for dEs reflect the vigorous
gravitational forces acting within the very environment
in which these galaxies typically reside.
Ram-pressure stripping \citep{gun72}, galaxy
harassment \citep{moo96}, and tidal stirring \citep{may01} are all
based on the removal of gas and the morphological transformation of a
late-type spiral or irregular galaxy, thereby attempting to reproduce
the seemingly plain appearance of dEs. On the other hand, differences
in the chemical abundances of early-type and late-type galaxies may
argue against a simple morphological transformation \citep{gre03}. In
any case, such structural transformations would be well-suited to explain the
famous morphology-density relation \citep{dre80}: the higher the
density is, the more efficiently are infalling spirals and irregulars
transformed into dEs, thereby skewing the relative
abundance of different types of galaxy towards massive early-type
objects as compared to abundances in the field.
 Moreover, \citet{con01} point out that the number of Virgo cluster dEs
 is more than a factor of 3 larger than what would be expected from just
 adding groups to the cluster. This strongly favours the idea that the
 majority of dEs were formed through a morphological transformation of
 galaxies that fell into the cluster.

Especially in recent years, small or intermediate-sized samples of early-type
dwarfs have been studied in a large variety of ways. \citet{bos05} find
the relation of far-UV--near-UV colour and luminosity to
behave opposite for early-type dwarfs and giants. \nocite{vZe04b}
Van Zee et al.\ (2004a) derive
intermediate ages and subsolar to solar metallicities for dEs via
optical multiband photometry. Similar values were reported by
\citet{geh03} from a Lick index analysis of high-resolution
spectra. These spectra and similar studies by \citet{vZe04a} and
\citet{sim02} also 
revealed a significant amount of rotation in some dEs. Finally,
\citet{buy05} presented HI $21\,{\rm cm}$ line observations
as a first study of the interstellar medium of a dE outside the Local
Group.

However, no formation scenario could yet be clearly confirmed or
rejected. This might be due to a very basic piece
of the puzzle still lacking: the unambiguous characterization of
early-type dwarf morphology.
Following common definition, early-type
dwarfs comprise both dwarf ellipticals and dwarf S0 (dS0)
galaxies -- 
we are not considering the fainter dwarf spheroidal
galaxies \citep[e.g.][]{gre03} or the ultra-compact dwarfs
\citep[e.g.][]{hil99} here.
The morphological appearance and overall profile of a dwarf elliptical
are clearly defined.
In contrast, dS0 galaxies are loosely defined as objects whose
overall appearance is similar to that of a dwarf elliptical, but where
a more detailed examination shows non-elliptical properties such as
lens shape or (central) asymmetries. \citet{bin91} argued that most of
these characteristics were indicative of a disk nature, and the authors
conjectured that ``many, if not most, dS0 systems must be disk
galaxies''. However, their existence as a separate
class of objects has been put in question by several authors
\citep[e.g.][]{ryd99}, and dS0s have frequently been 
treated as a subclass of dwarf ellipticals \citep[e.g.][]{bar03}.

The unambiguous discovery of disk substructure (spiral arms and/or
bars) in some dwarf ellipticals and dS0s
\citep{jer00a,bar02a,geh03,gra03,der03} eventually proved the presence of a
disk in at least some early-type dwarfs. At the same time, however,
this raised the question of whether these objects are genuine disk
galaxies,
 i.e.\ of flat oblate shape and without significant stellar spheroid,
or whether they are spheroids hosting
just a small disk component like the two low-luminosity ellipticals
presented by \citet{mor04}.
On the theoretical side, \citet{mas05} showed that a fraction of the
progenitor galaxy's disk is able to survive the morphological
transformation from galaxy harassment, providing a possible explanation
for disks in early-type dwarfs.

Since up to now, a systematic 
analysis of a large sample of early-type dwarfs for the presence of
disk features has been lacking, common practice has been to continue using
the original classification of the Virgo cluster catalog
\citep[VCC,][]{vcc}, therefore calling some 
objects 'dwarf elliptical', some 'dS0', and some 'dwarf elliptical with embedded disk'.
In order to avoid confusion we assign
the common abbreviation 'dE' to early-type dwarfs in general, thereby
meaning both dwarf ellipticals and dS0s. We shall then examine each
object for potential disk substructure, and introduce the term 'dEdi'
for a dE with disk features. 

Clearly, the small sample of dEdis discovered so far can neither serve
as basis for a revised classification nor is it sufficient to feed
formation theories with quantitative input concerning the fraction and
properties of such objects. A systematic search for disk
features in dEs is thus required, and is made possible by the Sloan
Digital Sky Survey (SDSS) Data Release 4 \citep[DR4,][]{sdssdr4}
which covers almost the whole Virgo cluster with multiband optical
imaging. With these data at hand, our study can properly address the
following questions: a) whether all objects listed as dS0 in the VCC indeed
show disk features, b) how large the fraction of galaxies with disk
features is among dEs, c) how this fraction is distributed with respect
to luminosity, d) where in the cluster these objects are located, and
e) whether they appear to be genuine disk galaxies, or just spheroids with a
disk component. The catalog of dEdis and dEdi candidates resulting from
this study will serve as important input for all future work on dEs,
since the observables under study (e.g.\ dE colours) can then be
correlated with the presence or absence of a disk.

Recently, \citet{agu05} have introduced a two-component 
definition of a dS0 based on one-dimensional profile fits, with those
(Coma cluster) objects
being called dS0s where a single S\'ersic fit did not lead to a satisfying
result and instead a combined S\'ersic plus exponential fit was
necessary. Our goal in this paper, in contrast, is to uncover disk
features on the two-dimensional image without any presumption
on one-dimensional profile shapes.  To investigate whether or not the
two definitions go hand in hand is beyond the scope of this paper, since it
requires that accurate profile fits be done for all our SDSS galaxies. This
will be the subject of a future paper in this series.

Our data and sample selection is described in Sect.\ \ref{sec:data}, followed
by an outline of the techniques for image analysis in Sect.\
\ref{sec:analysis}. Identifications of disk features are presented in
Sect.\ \ref{sec:results}. Section \ref{sec:spiral} focuses on
the quantitative measurement of spiral features. The flattening
distributions of the disk features and galaxies are analyzed in Sect.\
\ref{sec:flat}. The luminosity function and number fraction of dEs with
and without disk features is the subject of Sect.\ \ref{sec:magdist},
and the limitations in detecting disk features are considered in Sect.\
\ref{sec:limit}. In Sect.\ \ref{sec:spatial} we show how our objects
are spatially distributed within the Virgo cluster, and a discussion and
summary is given in Sect.\ \ref{sec:discuss}.


  \section{Data and sample selection}
 \label{sec:data}

  \subsection{SDSS images}
 \label{sec:sub_sdss}

  The SDSS DR4 covers
  all galaxies listed in the Virgo Cluster Catalog \citep[VCC,][]{vcc} with
  a declination of $\delta \lesssim 16\fdg25$, except for an
  approximately $2\arcdeg \times 2\fdg5$ area at $\alpha\approx 186\fdg2$, $\delta\approx
  +5\fdg0$ (see Fig.\ \ref{fig:distrnum}). It provides reduced and
  calibrated images taken in the 
  u, g, r, i, and z band with a pixel scale of $0\farcs396$, which
  corresponds to a physical size of $30\,\rm{pc}$ when adopting 
  $m-M=31\fm0$, i.e.\ $d=15.85\,\rm{Mpc}$. The SDSS imaging camera takes data in drift-scanning
  mode nearly simultaneously in five photometric bands, u, g, r, i,
  and z, and thus combines very homogeneous multicolour photometry
  with large area coverage, good resolution, and sufficient depth to
  enable a systematic analysis of early-type dwarfs.
  The images have an absolute astrometric accuracy of ${\rm RMS} \le
  0\farcs1$ per coordinate, and a relative accuracy between the r band
  and each of the other bands of less than $0.1$ pixels
  \citep{sdssastrometry}. They can thus easily be aligned using
  their astrometric calibration and need not be registered manually. The effective
  exposure time of $54\,\rm{s}$ leads for a bright dE ($\mb\approx 14$)
  to a typical total signal-to-noise ratio 
  (SNR) of about $1000$ in the r-band within an aperture radius of
  approximately two half-light radii. For a faint dE ($\mb \approx 18$) this value is
  typically about $50$.
  The RMS of the noise per pixel corresponds to a
  surface brightness of approximately $24.2\,{\rm mag/arcsec^2}$ in the
  u-band, $24.7$ in g, $24.4$ in r, $23.9$ in i, and $22.4$ in z.

  \subsection{Image stacking}
 \label{sec:sub_stack}
  
  In order to reach a higher SNR than that of the individual images, we
  produced a combined image by co-adding the g, r, and i-band
  images. The u and z-band images were not used, since their SNR is
  significantly lower and would thus lead to a decrease of the SNR of
  the combined image.
  When determining the sky level, proper object masks
  are required, so that pixels containing light from a star or a galaxy
  are excluded from the sky level calculation and only 'sky pixels'
  (i.e.\ pixels that contain nothing but sky background) remain
  unmasked. For this purpose, we 
  applied the Source Extractor Software \citep{sex} to each
  object's image and each band to yield a 'segmentation image'
  which marks the pixels of all detected sources by assigning them
  non-zero values. To ensure proper masking of all objects, we expanded
  the source areas on the segmentation image by smoothing it with a
  Gaussian filter, using \emph{IRAF}\footnote{IRAF is distributed by
  the National Optical Astronomy Observatories, which are operated by
  the Association of Universities for Research in Astronomy, Inc.,
  under cooperative agreement with the National Science Foundation.}
  \citep{iraf}. The resulting image serves as object mask.
  The
  sky level was then determined with \emph{IRAF\,/\,imstat} on the so-masked
  images along with the noise level, and was subtracted from the
  images. The g and i band images were shifted with 
  \emph{IRAF\,/\,imshift} to match the r band image; shifts were
  determined from the SDSS astrometry provided for each image (see
  above). We then applied weights $w_{\rm g,r,i}$ to each image,
  following \citet{kni04}:
  \begin{equation}
    w_{\rm g,i} = \frac{S_{\rm g,i}\,\sigma_{\rm r}^2}{S_{\rm
    r}\,\sigma_{\rm g,i}^2},\quad w_{\rm r} = 1,
  \end{equation}
  with $S_{\rm g,r,i}$ being the sky level and $\sigma_{\rm g,r,i}$ the
  noise level.
  The weighted g, r, and i-band images were then summed to form the
  final combined image for each object. The resulting total SNR is about a factor of
  $\sqrt{3}$ larger than in the r-band image.

  \subsection{Sample selection}
 \label{sec:sub_sample}

  From visual inspection of the
  combined images
  we chose a magnitude limit of $\mb=18\fm0$ for our study, with $\mb$
  provided by the VCC. This is the same magnitude limit up to which
  the VCC was found to be complete \citep{vcc}.
    Adopting $m-M=31\fm0$, it corresponds roughly to a limit in
  absolute magnitude of $\Mb\le -13\fm0$.
  A more thorough examination of our
  limitations in detecting disk features is presented later in
  Section \ref{sec:limit}.
  Initially, we selected all 552 cluster member and possible member
  galaxies with $\mb\le18\fm0$ that were classified as dwarf elliptical or 
  dS0 in the VCC, including those with uncertainties. We
  took into account the revised 
  membership and classification from \citet{virgokin}, as well as updated
  classifications for several objects given by
  \citet{bar02a,bar03}, \citet{geh03}, and \citet{lotz04}. 25 galaxies
  are not covered by the SDSS DR4. 25 objects
  with a classification 'dE/dIrr' were excluded, and also all the
  remaining objects were visually examined and excluded if they
  appeared to be possible dwarf irregulars due to asymmetric features
  in their image, which applied to 18
  galaxies. Thereby we avoid biasing
  our sample by the inclusion of potential non-early-type objects (which
  might be disk galaxies anyway). Three more objects (\object[VCC]{VCC0184},
  \object[VCC]{VCC0211}, \object[VCC]{VCC1941}) were classified as
  possible cluster members but appear to be 
  probable background spirals
because of their small size and their spiral arm structure,
  and were therefore excluded as well. Five more
  objects (\object[VCC]{VCC0615}, \object[VCC]{VCC0811}, \object[VCC]{VCC1052}, \object[VCC]{VCC1776}, \object[VCC]{VCC1884}) are of such low
  surface brightness that no examination for potential disk features
  is possible; these were also excluded. Our final sample comprises 476 early-type  
  dwarfs, 414 of which are definite members of the Virgo cluster according to
  \citet{vcc,virgokin}.


  \section{Image analysis techniques}
\label{sec:analysis}

  For bringing to light weak features that are hidden by the
  dominating and mostly smooth and symmetric overall light distribution, two
  methods have proven suitable. Unsharp masks are a common technique in
  detecing and enhancing weak substructure like e.g.\ nuclear bars or spirals
  \citep[e.g.][]{celeste,erw04}. They are produced by first smoothing
  an image and then dividing the original by the smoothed one, which can
  easily be performed automatically on a large dataset. Another technique is
  to model the smooth axisymmetric light distribution of a galaxy and
  subtract it from the original image \citep[e.g.][]{bar02a}, with
  non-axisymmetric features like spiral arms remaining. Both methods
  have been used to identify spiral arms, bars, or edge-on disks in
  eight Virgo cluster early-type dwarfs so far 
  (\object[VCC]{VCC0490}, \object[VCC]{VCC0856}, \object[VCC]{VCC0940},
  \object[VCC]{VCC1010}, \object[VCC]{VCC1036}, \object[VCC]{VCC1422},
  \object[VCC]{VCC1488}, \object[VCC]{VCC1695};
  \citealt{jer00a,jer01,bar02a,geh03,fer06}). These techniques are
  described below, along with the 
  derivation of an elliptical aperture for each galaxy, which is
  required as input for both methods.

    From our ongoing analysis of dEs with blue central regions (paper II of
    this series) we know that a significant fraction of dEs where no disk
    features were detected show obvious colour substructure.    
    Since we analyze the combined images from three bands in our search for
    disks, it could happen that colour substructure within the
    galaxy mimics the presence of a disk feature.
    To test this, we produced
    (uncalibrated) colour maps by dividing the aligned g and i-band
    images. Any detection of a disk feature with the methods outlined
    below can then be compared to the corresponding colour map and can
    thus be judged for reliability. To investigate
    whether or not there are any dEs in which colours do trace disk
    substructure requires a quantitative colour analysis that will be the
    subject of a future paper in this series.

  \subsection{Elliptical apertures}
 \label{sec:sub_ellipt}

  An elliptical aperture for each galaxy was determined by performing
  ellipse fits with \emph{IRAF\,/\,ellipse} on the combined image, allowing
  center, position angle, and ellipticity to vary. One of
  the outer elliptical isophotes -- usually between 1 and 2 half-light
  radii -- was then chosen by eye to trace best
  the \emph{outer} shape of each galaxy, as exemplified for VCC\,1010
  in the upper left panel of Fig.\ \ref{fig:vcc1010}. This ellipse was
  adopted to define the ellipticity and position angle of the galaxy.

  \subsection{Unsharp masks}
 \label{sec:sub_masks}
  
  We produced a set of unsharp masks for each object by smoothing the
  combined image with a two-dimensional circular and elliptical
  Gaussian, one at a time, of various kernel sizes $\sigma$. A small 
  value of $\sigma$ will enhance small structures and weaken large
  features at the same time, while a large kernel size will enhance large
  structures over small ones. For each set of unsharp masks we chose values of
  $\sigma=2,4,6,9,13,20,$ and $30$ pixels.
  With {\bf $d=15.85\,\rm{Mpc}$} ($m-M=31\fm0$) and a
  subsequent pixel scale of $77\,{\rm pc/arcsec}$ ($30\,{\rm pc/pixel}$),
  these values correspond to $0.06, 0.12, 0.18, 0.27, 0.40, 0.61,$ and
  $0.91$\,kpc, respectively.

  It is desirable to produce both masks created with a
  circular Gaussian (hereafter referred to as 'circular masks') and
  masks with an elliptical Gaussian ('elliptical masks') corresponding to the
  galaxy's ellipticity and position angle. Circular masks of
  non-circular artificial galaxies show a characteristic narrow shape along
  the major axis that could easily be confused with an edge-on
  disk and does not occur when applying elliptical masks. We
  demonstrate this in Fig.\ \ref{fig:simul}, where a dE is represented
  by a two-dimensional exponential surface brightness profile with an
  elliptical shape created with \emph{IRAF\,/\,mkobjects} (left panel).
  A circular unsharp mask with a 
  Gaussian kernel of $\sigma=4\,\rm{pix}$, feigning an edge-on disk, is
  shown in the middle panel. In the right panel, an elliptical mask
  with position angle and ellipticity matching that of the galaxy has
  been applied: no substructure is seen. This is due to the fact that
  the scale radius of the light profile is smaller along the minor
  axis; therefore an isotropic Gaussian will blur the object much
  stronger along the minor than along the major axis. For
  detection of egde-on disk features or bars that are roughly parallel to the
  major axis, elliptical masks are thus clearly preferred. However,
  frequently the inner isophotes of an object
  are significantly rounder than the outer ones that define
  the Gaussian's ellipticity. In these cases, again an
  artificial narrow (bar-like) structure will appear along the
  \emph{minor} axis, due to the very same effect as described
  above. Here, circular masks serve as a complementary check whether an
  apparent elongated feature along the minor axis is real or is only caused
  by varying ellipticity.

  \subsection{Residual images from ellipse fits}
 \label{sec:sub_residual}
  
\begin{figure}
  \epsscale{0.45}
  \plotone{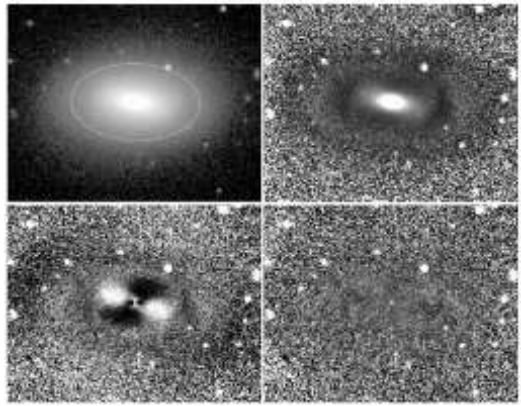}
  \caption{{\bf Image analysis techniques.}
    {\it Upper left panel:} Combined
    image of VCC1010, along with the elliptical isophote defining its
    shape.
    {\it Upper right panel:} Elliptical unsharp mask with
    kernel size $\sigma=20\,\rm{pix}$.
    {\it Lower left panel:}
    'Fixed model' residual image, i.e.\ produced via ellipse fits with
    fixed ellipticity and position angle.
    {\it Lower right panel:} 'Variable model'
    residual image, i.e.\ produced via ellipse fits with variable
    ellipticity and position angle.
      Each panel has a horizontal scale of $300$ pixels ($119''$ or $9.13\,$kpc with
    $d=15.85\,\rm{Mpc}$, i.e.\ $m-M=31\fm0$).
  }
  \label{fig:vcc1010}
\end{figure}

  A galaxy's surface brightness distribution can be modeled by
  performing ellipse fits (with \emph{IRAF\,/\,ellipse}) and then
  feeding the output directly into the task \emph{bmodel}. The
  resulting model image is then subtracted from the original object,
  yielding a residual image. Any information contained in 
  the results of ellipse fitting directly enters the model. This can
  nicely be demonstrated on \object[VCC]{VCC1010}, which hosts a
  bar. If we construct a model through ellipse fits with \emph{variable}
  position angle and ellipticity, the bar
  is not seen at all in the residual image (lower right panel of Fig.\
  \ref{fig:vcc1010}) since it has
  been fully reproduced by the model. If position angle and ellipticity
  are instead fixed at a value taken well outside the bar (namely the
  chosen elliptical aperture as described above),
  a strong residual double-cone is seen (lower left panel of Fig.\
  \ref{fig:vcc1010}), which has already been explained by \citet{bar02a}
  as characteristic shape of a changing position angle, and therefore
  of a bar. Similarly, spiral arms can be reproduced to a large extent
  by varying ellipses, and thus do not appear in the residual image
  unless position angle and ellipticity are kept fixed.
  
  From the above considerations it is obvious that any disk feature can
  best be detected with a model built through fixed ellipticity and position
  angle (later referred to as 'fixed model'). However, in principle any
  \emph{additional} weak, asymmetric features would require
  \emph{variable} ellipse parameters ('variable model'), 
  so that the bar or spiral is properly reproduced in the model and
  fully subtracted from the image, and the additional substructure
  remains. Therefore, both types of residual images were visually
  examined along with the unsharp masks for each object.

  \subsection{Artificial galaxies}
 \label{sec:sub_simul}

  In addition to the SDSS data we produced artificial dE
  galaxies with \emph{IRAF\,/\,mkobjects}, adopting a two-dimensional
  exponential surface brightness profile with an elliptical shape (left
  panel of Fig.\ \ref{fig:simul}). This 'primary' object was then
  superposed by another 'secondary' exponential light distribution with
  the same or higher ellipticity, representing an (inclined) disk
  within a spheroid (Fig.\ \ref{fig:twocomp}). Various
  primary-to-secondary flux ratios, scale ratios, position angles and
  inclinations were reproduced, in order to provide a model counterpart for
  real galaxies that potentially are spheroids
  hosting a disk. The noise characteristics of the artificial images
  were chosen to be similar to a typical SDSS image, and galaxies
  covering a range of SNR values were created.


   \section{Results: early-type dwarfs with disk features}
 \label{sec:results}

   Close visual inspection of the combined image, the set of unsharp masks,
   and the two residual images was performed for each galaxy, using the
   \emph{SAOImage DS9} tool \citep{ds9}. It turned out
   that unsharp masks are the primary means to search for substructure:
   especially for small elongated features, they often provide a more reliable
   and clearer detection than the residual images do. In turn, only in very
   few cases did the residual images show hints of substructure where the
   unsharp masks did not. However, in these cases the features were weak
   and their shape hard to define.
   Therefore we adopted a conservative approach and did not consider them
   as possible substructure.
   As \citet{bar03} pointed out, care must be 
   taken with features seen solely on the residual images, 
   since the models can be deceived by e.g.\ changing ellipticity and position
   angle, so that the resulting residual image would feign some
   substructure where none is present. Furthermore, the variable model
   turned out to be of little use, since it either reproduces substructure
   completely and yields a blank residual image (see Fig.\
   \ref{fig:vcc1010}), or leaves only weak features that are readily seen in
   the unsharp masks and the fixed model residual image. The situation
   described above that the variable model would bring to light secondary
   features by reproducing and subtracting the primary ones did not occur,
   i.e.\ no secondary substructure remained in the residual image other than
   weak and highly doubtful features.

   \subsection{Disk detections}
 \label{sec:sub_disks}

\begin{figure}
  \epsscale{0.45}
  \plotone{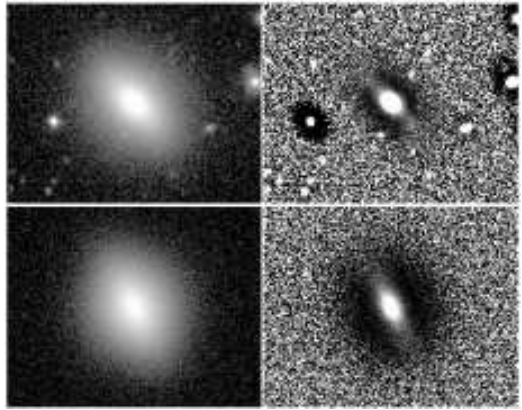}
  \caption{{\bf Simulated vs.\ observed dEdi.}
    {\it Upper panels:} Combined image of VCC0990 along with its 
    elliptical unsharp mask ($\sigma=9\,\rm{pix}$).
    {\it Lower panels:} Simulated
    two-component galaxy image along with its elliptical unsharp mask
    ($\sigma=9\,\rm{pix}$). The 'primary' component 
    has an exponential intensity profile with scale length 30 pixels, axial ratio
    1. The 'secondary' component has an exponential intensity profile with
    equal scale length, axial ratio 0.5, and a total magnitude $0\fm5$
    fainter than that of the primary component. The
    parameters are chosen to roughly match the appearance of VCC0990. Note
    that the simulation contains no nucleus, which is why the central region
    of the unsharp mask is brighter in the observed image than in the
    simulated one.
      Each panel has a horizontal scale of 248 pixels ($98''$ or $7.55\,$kpc with
    $d=15.85\,\rm{Mpc}$).
  }
  \label{fig:twocomp}
\end{figure}

   We identified 14 out of 476 early-type dwarfs that unambiguously show
   disk features, as exemplified in the upper three panels of Fig.\
   \ref{fig:pics}.
   Moreover, we find 'probable disks' in 10 objects (third
   panel from bottom of Fig.\ 
   \ref{fig:pics}), and 'possible disks' in 17 objects (lower two panels of
   Fig.\ \ref{fig:pics}).
   This distinction between 'unambiguous',
   'probable', and 'possible' disks is based on the visual judgement of all
   three authors, and is intended to be an honest representation of the
   (un)ambiguity and the SNR of disk features. In the case of a possible
   edge-on or inclined disk, we used comparisons with artificial
   two-component galaxies to check whether our
   interpretation is consistent with such a structure. This is exemplified in
   Fig.\ \ref{fig:twocomp}, where the galaxy \object[VCC]{VCC0990} -- classified as
   'probable' dEdi -- is compared to an artificial galaxy consisting of a
   'primary' and a 'secondary' component, the latter being fainter and having
   a larger ellipticity (i.e.\ representing a larger inclination angle). The
   simulated image is chosen to be similar in SNR and size, and indeed the
   shape of the galaxy images as well as their unsharp masks look similar.

     In two cases (\object[VCC]{VCC1684} and \object[VCC]{VCC1779}),
     the colour maps (see Sect.~\ref{sec:analysis}) show a blue central
     region that is similar in appearance to the possible disk
     features. As a further test we produced unsharp masks for the two
     galaxies from the i-band images only. However, in both cases we
     can neither reject nor unambiguously confirm the presence
     of an inclined disk. We thus list both objects as showing
     'possible disk' features.

   In several cases we could
   not decide whether we see an edge-on disk or a bar; nevertheless, both were
   taken as disk feature, since the presence of a bar commonly requires a
   disk. Moreover, apart from the simple category 'no substructure detected'
   (applying to 406 objects listed in Appendix \ref{sec:nodisk}), we
   labelled 29 galaxies as objects where 
   substructure of some kind 
   is present, but not necessarily indicative of a disk
   ('other
   substructure'; objects listed in Appendix \ref{sec:nodisk}).
     17 of these show irregular central features (also see Sect.\
   \ref{sec:sub_dS0}), five have a boxy shape, in four objects a
   feature like a dust lane is seen, and for three objects the unsharp
   masks appear to show a luminosity excess in the inner part.

   Of the eight Virgo dEs for which disk features have been reported,
   five (\object[VCC]{VCC0490}, \object[VCC]{VCC0856},
   \object[VCC]{VCC1010}, \object[VCC]{VCC1036},
   \object[VCC]{VCC1695}) are contained in 
   our 14 unambiguous detections, and one (\object[VCC]{VCC1422}) is a probable
   detection. Both \object[VCC]{VCC0940} \citep[reported by][]{bar02a} and \object[VCC]{VCC1488}
   \citep[reported by][]{geh03} were not even identified as a
   dEdi candidate by us. The reason might be twofold: first, those
   studies \citep[as well as][]{der03} use a boxcar or median filter to
   create their unsharp masks. As we
   demonstrated above (see Fig.\ \ref{fig:simul} and Sect.\
   \ref{sec:sub_masks}), applying such a filter to a perfectly smooth
   elliptical light distribution will yield an artificial elongated
   structure in the unsharp mask. This effect might well apply to
   \object[VCC]{VCC1488} with its axial ratio of 0.55, but less likely to \object[VCC]{VCC0940}
   which has an axial ratio of 0.76. However, the disk features of
   both galaxies were also reported to be seen in the
   residual images resulting from ellipse fits and subsequent modeling
   of the light distribution. Given that both the data from
   \citet{geh03} and from \citet{bar02a} are of higher depth and
   resolution than our SDSS images, the non-detection of
   ours might simply reflect our limitations in detecting disks, and
   shows that more dEdis might exist than those identified by us (see also
   Sects.\ \ref{sec:magdist} and \ref{sec:limit}). 
   
   We list the dEdis and dEdi candidates in Table \ref{tab:disks}. We do,
   however, not attempt to reclassify objects, since classification
   schemes in the VCC were fairly complex and based on the surface
   brightness distribution, whereas we aim solely at stating whether or not a dE's image
   shows features of a disk. In principle, it would be desirable to establish a
   'pure' definition of the dS0 class as those (and only those) dEs
   hosting (or being) a disk. Unfortunately, this is not possible: apart
   from the fact that many objects can only be termed candidates due to
   the limited SNR, those where no disk was \emph{found} do not
   necessarily have to \emph{have} no disk. It appears therefore most
   useful to not touch the original VCC classification, but instead 
   to provide a list of (candidate) dEdis that can be correlated with all sorts of
   observables in future studies of dEs. A thorough reclassification of
   all galaxies is deferred to a future study. We point out that our
   objects are \emph{not} related to the so-called dwarf spiral
   galaxies defined by \citet{schom95}: while those have a classical
   bulge, our objects do not.
  
   \subsection{Correlation with the original dS0 class}
 \label{sec:sub_dS0}
   
   \citet{bin91} described five cases in which a galaxy was classified
   dS0, with characteristics mostly indicative of a disk nature of
   the galaxy. Briefly, criteria for dS0s were a bulge-disk-like
   profile, high flattening, a lens-like appearance, a global asymmetry
   (like a bar or boxiness), and an irregularity in the central
   part.

\begin{figure}
  \epsscale{0.45}
  \plotone{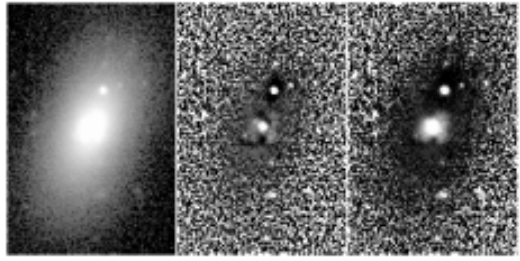}
  \caption{{\bf A dE with irregular central substructure.}
    Combined
  image of VCC0781 (left panel), unsharp mask image with kernel size
  $\sigma=4\,\rm{pix}$ (middle panel), and unsharp mask with
  $\sigma=9\,\rm{pix}$ (right panel). Of those dEs where
  substructure other than disk features was found, this galaxy
  represents the subgroup of objects with central irregularities likely to be
  caused by gas and/or dust.
      Each panel has a horizontal scale of $46''$ ($3.53\,$kpc  with
    $d=15.85\,\rm{Mpc}$).
  }
  \label{fig:E8}
\end{figure}

   Our initial sample -- prior to exclusion of possibly irregular
   objects -- contained 47 out of 50 galaxies
   classified as dS0 or candidate dS0 (e.g.\ 'dE or dS0') in the
   VCC. Two objects were then exluded due to a possible irregular
   nature; thus 45 (candidate) dS0s are left in our working sample.
     22
   of these are indeed classified by us as dEdis or dEdi
   candidates, constituting 54\% of our dEdi sample.
   14 objects
   have 'other substructure'
   which reflects the criteria of
   \citet{bin91}: 3 of them have a boxy shape, and 9 show irregular or
   clumpy central features likely caused by gas and dust. As an example
   for the latter, we show in Fig.\ \ref{fig:E8} the image and unsharp
   masks of \object[VCC]{VCC0781}, which looks somewhat similar to the well-known
   dwarf elliptical NGC205 in the Local Group.
   Interestingly, all of
   these 9 objects
   with central gas/dust  
   features have a blue central 
   region
   with ongoing star formation or at least very young
   stars, similar to NGC205 and also to the galaxy presented by \citet{gu06}. This nicely
   confirms Binggeli's \& Cameron's conclusion, ``the
   irregularity must stem from recent or ongoing star formation''
   (drawn without colour information or unsharp masks!). None of these galaxies
   shows (additional) disk features; thus caution must be taken when
   treating them as dEdis only because of their dS0 class: not all
   classified dS0s are dEdis.
   These objects
   might prove highly important for investigating possible formation
   channels for dEs; therefore they will be the subject of paper II of
   this series (Lisker et al., in prep.). 

   Finally, for 9 of the 45 (candidate) dS0s, neither a disk nor other substructure
   was found. However, three of these are classified 'dE or dS0', three are
   'dS0?' (i.e.\ high uncertainty),
     and two are 'dS0:' (i.e.\ some uncertainty); hence we most
   probably did not miss any significant disk or irregular substructure.
     The one unambiguously classified dS0 (\object[VCC]{VCC1912}) had been classified as
   such mainly due to high flattening. While our measured axial ratio of 0.33
   is small, it is not small enough that we would classify it as dEdi
   based on flattening only.


   \section{Properties of spiral features}
 \label{sec:spiral}

   \subsection{Relative strength}
 \label{sec:sub_strength}
   
   For those three dEdis with the best-defined spiral arms, we now
   attempt to obtain an estimate of the relative amount of light that
   constitutes the spiral arms, as compared to the smooth and axisymmetrically
   distributed light. We thus need to measure the flux of the residual image
   (showing only the spiral arms) within a given aperture, and compare it to
   the total flux of the galaxy within the same aperture. We shall
   term this flux ratio the 'strength' of the spiral features. However,
   in the residual image the flux level in between the spiral arms is
   significantly negative: when fitting ellipses, the average flux value of
   each elliptical isophote is affected by the spiral arms and thus comes out
   slightly too high. Consequently,
   somewhat too much flux is assigned to the smoothly distributed light
   component, resulting in negative flux values when subtracted from the
   original image. To avoid or at least minimize this effect, we obtain
   optimized residual images through an iterative procedure outlined in
   detail in Appendix \ref{sec:resiopti}, yielding a lower and an upper
   limit for the strength of the residual features.

   The resulting residual images for our three dEdis are
   presented in Fig.\ \ref{fig:strength}.
   Note that it is \emph{not} the case that our disk detections would have
   been more efficient if we had used such optimized residual images from the
   beginning: the \emph{contrast} of residual features like spiral arms does
   not differ with respect to the initial residual images -- only the average
   flux level is offset systematically.

   Apertures enclosing the spiral arms were now chosen manually, and the strength of the
   spirals was measured from the residual and the model flux within the
   same aperture. The nucleus as well as foreground stars or
   background objects were masked to avoid any bias. The results are listed in
   Table \ref{tab:strength}: \object[VCC]{VCC0490} has the strongest spiral features, which
   amount to 12-13\% of the total light. The spiral of \object[VCC]{VCC0308} constitutes
   8-11\% of the light, and \object[VCC]{VCC0856} only reaches 6-8\%. 

   With these results at hand, we can now for the first time in the course of
   this paper address the
   question of whether dEdis are disk galaxies, i.e.\ are of flat oblate
   shape like \object[VCC]{VCC1304} (third row in Fig.\
   \ref{fig:pics}), or whether they are spheroids hosting a disk
   component. The ratio of the light within the spiral features to the
   smoothly distributed light has been measured to be within 6-12\% for
   our three galaxies. Therefore, when assuming that these objects are
   spheroidal galaxies hosting an embedded disk, the total light within
   the disk cannot be much larger than the light within the spiral
   features, since otherwise the disk would be the dominating component
   and the object would not be a spheroidal galaxy in the common
   sense. Therefore, assuming the light within the spiral features to
   be of the same order as the total light of the disk component,
   the above ratio of 'spiral light' to the smoothly distributed light
   should be comparable to the ratio of the secondary to the primary component
   in our two-component model images.
   If, however, our galaxies would be genuine disk galaxies, the spiral
   features might well contain 
   just a fraction of the total light of the disk. Consequently, if the
   disk is seen edge-on and compared to a suitable two-component model image,
   the ratio of its secondary to primary component should be
   significantly larger than the value measured for the (face-on)
   spiral features. Indeed, for those dEdis with apparent inclined
   disks that could not be confused with a bar, the
   secondary component of the similar-looking model images is only
   0.5-1 magnitudes fainter than the primary component, whereas the
   spirals measured above are 2.2-3.1 
   magnitudes fainter than the smooth axisymmetric component. Although
   this is no
   final proof due to the small number of objects considered, it
   points towards dEdis \emph{being} disk galaxies, instead of just
   \emph{having} a disk component. 
Further arguments supporting this
   view will be presented in Sect.\ \ref{sec:flat}.

   \subsection{A possible connection to faint S0/Sa galaxies}
 \label{sec:sub_Sa}
   
   Since the strengths of the three spirals measured above already
   differ within a factor of 
   two, it might be interesting to see how the galaxies' images would appear
   if their spirals were stronger by a certain amount. For this purpose, we
   simply multiplied the residual images by a certain factor
   and added them to the model of the smooth component, thereby mimicking a
   stronger spiral. Strikingly, with only a 0.5-1\,mag enhancement,
   the galaxy does not look like a dwarf
   elliptical or dwarf S0 anymore,
 but instead like a spiral galaxy, although
   without a bulge.

\begin{figure}
  \epsscale{0.45}
  \plotone{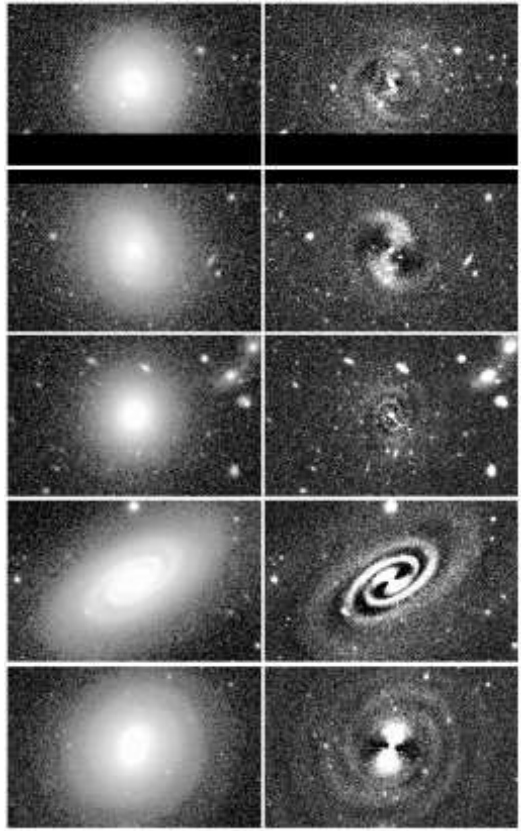}
  \caption{{\bf Residual images of spiral arms.} 
    Combined images as well as optimized residual images as described
    in Sect.\ \ref{sec:sub_strength} are shown for the three dEdis with
    the best-defined spiral structure (VCC0308, VCC0490, and VCC0856
    from top), as well as for the two dwarf-like S0/Sa galaxies (Sect.\
    \ref{sec:sub_Sa}) VCC0522 and VCC1902 (bottom).
      Each panel has a horizontal scale of $162''$ ($12.48\,$kpc  with
    $d=15.85\,\rm{Mpc}$).
  }
  \label{fig:strength}
\end{figure}

 It
   might thus be no coincidence that more than a 
   decade ago, one of us (B.B.) identified a handful of ``faint,
   dwarfish looking S0/Sa'' galaxies
   in the Virgo cluster
   (\object[VCC]{VCC0522}, \object[VCC]{VCC1326},
   \object[VCC]{VCC1368}, \object[VCC]{VCC1757}, \object[VCC]{VCC1902}) 
 whose appearance is very similar to what has been just
   described (Fig.\ \ref{fig:strength}). These objects differ from
   normal (i.e.\ giant) S0/Sa galaxies: their surface 
   brightness profiles are similar to early-type dwarfs and remain
   flatter than the flattest possible King profile when going inwards,
   i.e.\ they apparently have no bulge 
(Binggeli, unpublished). 
Thus, they are hardly normal
   S0/Sa galaxies, which typically have a high bulge-to-disk ratio. Instead they have a central luminosity excess just like the early-type
   dwarfs.

One might thus term these objects 'dwarf-like S0/Sa' galaxies, to
   distinguish them from their giant counterparts. A further
   investigation of their characteristics and a detailed comparison with
   early-type dwarfs will be the subject of a future paper in this series.
 For our present
   study, we selected those two with the best-defined spiral
   structure (\object[VCC]{VCC0522}, classified Sa, and \object[VCC]{VCC1902}, classified S0/Sa),
   in order to measure the spiral strength like we did above 
   and compare it to the dEdis. Their strengths turn out to be slightly
   larger than the average value of the 
   three dEdis and similar to the strongest dEdi spiral (\object[VCC]{VCC0490}): 9\%-13\%
   for \object[VCC]{VCC1902} and 11-14\% for \object[VCC]{VCC0522}. Both objects are about half a
   magnitude brighter than the brightest dE(di)s. It thus appears
   plausible that the dEdis and these objects belong to the same
   population of galaxies that extends to magnitudes brighter than
   those of dEs and differs from the 'classical' dwarf ellipticals.

   \subsection{Pitch angle}
 \label{sec:sub_pitch}
   In order to confirm our above hypothesis, we measured the pitch angle of
   the spiral arms of both dEdis and the dwarf-like S0/Sa galaxies on
   the residual images. We 
   used the method described by \citet{ma01}: a spiral arm is traced by
   manually selecting a series of image positions that follow the arm. These
   are then fitted by a logarithmic spiral, taking into
   account the galaxy's inclination and position angle which we adopt from our
   elliptical apertures (in the case of \object[VCC]{VCC1896} these values were taken
   from the axial ratio measurement of the disk). We measured two arms of \object[VCC]{VCC0308},
   \object[VCC]{VCC0490}, \object[VCC]{VCC0856}, and \object[VCC]{VCC1896}, one arm of each of the two possible
   cluster members \object[VCC]{VCC0278} and \object[VCC]{VCC1671}, and two arms of the two faint
   S0/Sa galaxies \object[VCC]{VCC0522}
   and \object[VCC]{VCC1902}. The
   resulting values are shown in Fig.\ \ref{fig:pitch} as black symbols and
   compared to the values for various Hubble types from \citet{ma99}
   (grey). The dwarf-like Sa \object[VCC]{VCC0522} falls within the range of values of
   the dEdis while the dwarf-like S0/Sa \object[VCC]{VCC1902} lies slightly below. The
   dEdis best agree with Hubble type Sab/Sb while the dwarf-like S0/Sa
   galaxies -- if taken together -- fall in the range of type Sab. An
   independent check of our measurements is provided by \citet{jer00a}
   who find a pitch angle of $12\fdg1$ for \object[VCC]{VCC0856}. For the two arms,
   we derive the values $10\fdg5$ and $12\fdg1$, respectively, thus
   being in good agreement with those VLT-data measurements. Our derived
   pitch angles are incompatible 
   with spirals of very late type ($>$Sc), which are often considered as potential
   progenitors for dEs; see Sect.\ \ref{sec:discuss} for a discussion.


   \section{Flattening distribution}
 \label{sec:flat}

   A flattening distribution for our galaxies can be obtained in two
   ways: for the disk features themselves by directly measuring
   or estimating their axial ratio, and for the galaxies as a whole,
   based on their ellipticities. The first distribution -- which we
   shall term the flattening distribution of the disks -- serves as a
   basic test that the features we see are indeed disk
   features. This is of particular importance for the inclusion of
   'probable' and 'possible' disk features into our working sample of
   dEdis. In order to have a statistically significant sample, e.g.\ to
   derive the luminosity function (Sect.\ \ref{sec:magdist}), we would
   like to include not only those dEs with unambiguous disk features,
   but also those with probable and possible disk features into our
   dEdi working sample. This requires the flattening distribution of
   disk features to be consistent with the assumption of an intrinsic
   flat oblate (and circular) shape, which shall be examined in the following subsection.

   The flattening distribution of the \emph{galaxies} -- presented below in
   Sect.\ \ref{sec:sub_flatgal} -- serves a different
   purpose: it will allow us to consider the question of the
   possible disk nature of the dEdis again. If they were spheroidal galaxies
   with a (weak) disk component, the distribution of axial ratios
   should be significantly different from that of disk galaxies. In
   turn, if their flattening distribution would be consistent with them
   having an intrinsic disk shape, they would very likely be
   genuine disk galaxies.

   \subsection{Flattening distribution of the disks}
 \label{sec:sub_flatdisk}
   
\begin{figure}
  \epsscale{0.45}
  \plotone{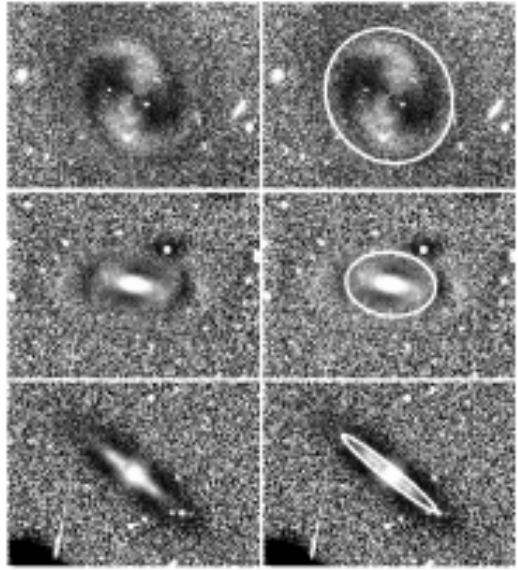}
  \caption{{\bf Disk axial ratio measurement.}
    Illustration of manual
  choice of a best-fitting elliptical aperture (right panels) for each
  disk feature. From top to bottom: VCC0490 (residual image), VCC1010
  (unsharp mask with kernel size $\sigma=13\,\rm{pix}$), and VCC1304
  (unsharp mask with $\sigma=20\,\rm{pix}$).
      Each panel has a horizontal scale of $116''$ ($8.95\,$kpc with
    $d=15.85\,\rm{Mpc}$).
  }
  \label{fig:incl}
\end{figure}

   Although not possible with perfect accuracy, still an estimate
   of the inclinations of the disks (not the galaxies) can be obtained from
   either the unsharp mask or the residual images. An ellipse was
   manually (by eye) fitted to the disk using that unsharp mask or
   residual image where the respective features stand out most
   prominently (exemplified in Fig.\ \ref{fig:incl}). The results are
   shown in the left panel of Fig.\ \ref{fig:axgaldisk} as a running
   histogram (black lines), i.e.\
   at each data point we consider the number of objects within the
   chosen bin-width of 0.1 ($\pm$0.05). We take into account all
   36 dEdis and candidates that are certain cluster members. Galaxies
   where we 
   cannot decide whether we see a bar or an edge-on disk were assigned two
   values: a lower limit assuming an inclined disk (solid
   line), and an upper limit from the axial ratio of the galaxy as a
   whole, assuming the feature was a bar (dashed line). A theoretical
   distribution assuming a disk with an intrinsic axial ratio following
   a narrow Gaussian around a mean value $\mu=0.25$ with $\sigma=0.01$ and a
   randomly distributed inclination is shown as grey solid line for
   comparison \citep{mihalasbinney}. Within the expected
   uncertainties for our relatively 
   crude measurements, the observed and theoretical curve are nicely
   consistent with each other. This strongly corroborates the
   hypothesis that the features we see
   \emph{are} disks, and moreover, it supports the approach of
   including not only the unambiguous but also the candidate objects
   into our dEdi working sample for the purposes of our analysis.
   As a further test, we examined the flattening distribution for
   'possible disks' only -- it turns out to be very similar to the
   distribution for
   all dEdis. It therefore seems plausible that most of our 'possible' disk
   detections actually are disks. Nevertheless we prefer to keep the
   term 'possible' in order to reflect that uncertainties \emph{are}
   present in our visual identification of disk features.

   \subsection{Flattening distribution of the galaxies}
 \label{sec:sub_flatgal}

\begin{figure}
  \epsscale{0.45}
  \plotone{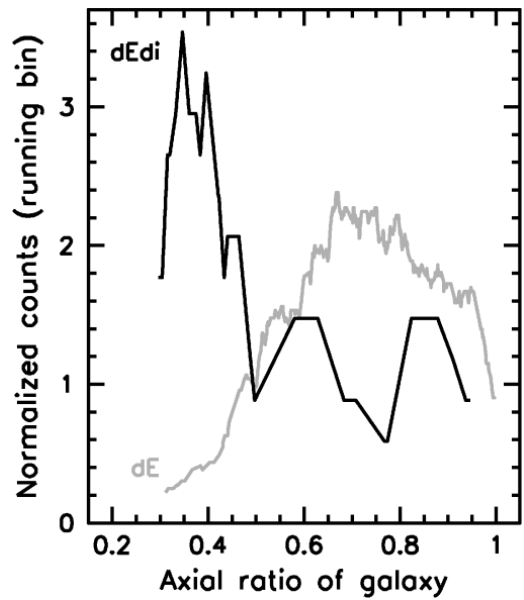}
  \caption{{\bf Galaxy axial ratio distribution.}
    Running histogram of the galaxy axial ratio distribution of dEdis
    (black) and dEs where no disk features were found (grey). Both
    histograms are normalized to an area of 1.
  }
  \label{fig:axgalcomp}
\end{figure}

   Based on the elliptical apertures described in Sect.\
   \ref{sec:sub_ellipt} we put together the distribution of axial
   ratios of the (candidate) dEdis, shown in the right panel of Fig.\
   \ref{fig:axgaldisk} as running histogram (black line). For
   comparison, we show the theoretical curve assuming an intrinsic
   axial ratio distribution given by a narrow Gaussian with $\mu=0.35$
   and $\sigma=0.02$. Obviously, there is almost perfect agreement of
   observed and theoretical distribution, a compelling indication for
   an intrinsic disk nature of the dEdis! This view gains further
   support from the comparison with the distribution of dEs where no
   disk features were found (Fig.\ \ref{fig:axgalcomp}): these objects
   are clearly consistent with a population of spheroids, and differ
   significantly from the dEdi distribution. It thus appears very
   likely that dEdis are genuine disk galaxies. A prototypical
   representation of how these disk galaxies appear when
   viewed egde-on might be given by \object[VCC]{VCC1304} (third row in Fig.\
   \ref{fig:pics}) with its axial ratio of 0.32. 

While \citet{bin95} already found dS0s to be significantly flatter than
  dwarf ellipticals, the difference is even more pronounced for our
   comparison of dEdis and dEs with no disk detection. This is explained
   by the fact that not all dS0s are dEdis and vice versa: at least
   some galaxies that were classified as dS0 might be spheroids (see
   Sect.\ \ref{sec:sub_dS0}).

   The flattening distribution also allows us to test whether or not
   \emph{all} bright dEs might actually be dEdis, but are not
   identified as such due to limitations of our detection method. When
   we modify Fig.\ \ref{fig:axgalcomp} such that only galaxies of the
   brightest one-(two-)magnitude-interval are considered (not shown), the
   distribution of dEs with no disk detection is inconsistent with all
   of them
   being dEdis as well. Therefore, while we might miss \emph{some}
   dEdis in our search for disk features as outlined in Sect.\
   \ref{sec:limit}, we can exclude the possibility that \emph{all} of
   the brightest dEs are disk galaxies -- a significant number of
   objects need to be spheroids.


   \section{Disk fraction versus magnitude}
 \label{sec:magdist}

   In the upper panel of Fig.\ \ref{fig:maghist} we show the
   distribution of dEs and (candidate) dEdis 
   with respect to their B band magnitude provided by the
   VCC. For this purpose we present our data as a running histogram
   with a bin-width of 1\fm0 (i.e.\ $\pm$0\fm5). Only galaxies are
   considered that are certain cluster 
   members according to \citet{vcc,virgokin}, resulting in 414 objects
   (light grey shaded histogram), containing 36 dEdis and candidates (dark grey
   shaded). The fraction of (candidate) dEdis among all dEs is shown as
   black solid line: it reaches more than 50\% for the brightest
   objects, and then decreases to few percent at $\mb>16$.
   This 'disk fraction' might be of special interest, since e.g.\ \citet{bin91}
   discuss a potential break in dwarf galaxy structure at
   $M_{B_T}\simeq -16$, which corresponds to $\mb=15.7$ given their
   $m-M=31.7$.

   A plateau is seen in the running histogram (the luminosity function)
   of our full dE
   sample, the position of which coincides very well with the
   location of the dEdis in the diagram. As a test, we subtracted the
   dEdi-counts from those of the full sample, but still a weak bump
   remains. However, we need to take into account the
   fact that we might have missed a significant number of disks in dEs
   due to the limitations of our data (which are assessed in Sect.\
   \ref{sec:limit}). Therefore, we now multiplied the dEdi counts with
   1.5 to account for the missed ones, and subtracted these counts 
   from those of the full sample. Indeed, the plateau disappears (lower
   panel of Fig.\ \ref{fig:maghist}).

   These results -- independent of any considerations in previous
   sections -- suggest very convincingly that
   dEdis are a different population than dEs with no disk, i.e.\ both
   have different origins not related to each other. Taken together
   with the indications for the disk nature of dEdis, evidence
   accumulates that dEdis are not just dwarf ellipticals with embedded
   disks, but instead constitute a population of disk galaxies
   different and independent from classical dwarf ellipticals.

   In the following, we attempt to estimate the number of disks that
   are missed by our study, in order to assess whether the above
   assumption of a factor of 1.5 is realistic. Moreover, we attempt to
   independently show that the decline of the disk fraction is real,
   and cannot be just an effect of limited data quality.


   \section{Limitations in detecting disks}
 \label{sec:limit}
   
   To obtain a realistic estimate for the limitations in detecting
   disks, we artificially dimmed our objects such that they
   correspond to dEs that are fainter by one (two) magnitude(s), also
   taking into account the relation of dE magnitude and radius
   \citep{bin91}. This was done on the individual images and is
   described in more detail in Appendix \ref{sec:dim}. The resulting
   modified images were then coadded like the original data, and
   unsharp masks were created. The dimmed objects were then
   treated as if they were real galaxies that have to be searched
   for disk features, and the same categories ('unambiguous',
   'probable' etc.) were assigned.

   In Fig.\ \ref{fig:magfaint1}, we focus on the galaxies
   lying within the brightest one-magnitude-interval (solid line, filled
   circles). When dimmed by 1 magnitude, they result in the histogram
   given by the dashed line with crosses,
   and when dimmed by two magnitudes, the resulting histogram is shown
   by the dot-dashed line with filled triangles. This is illustrating
   the disk fraction we would \emph{expect} to see at fainter
   magnitudes \emph{if} the fraction of the brightest one-magnitude-interval of our
   sample would be constant with
   magnitude\footnote{Here we neglect the fact that there is already a
   large decrease in disk fraction \emph{within} the brightest
   one-magnitude-interval -- however, a certain interval width is necessary in
   order to still have a fair number of dEdis left among the
   two-magnitude-dimmed dEs.}.
   The obvious mismatch, along with the already strong decrease in disk
   fraction \emph{within} the brightest one-magnitude-interval itself, suggests
   that the observed decline in disk fraction is real, and is not due
   to the limitations of the data. Even if we do not assume the true
   fraction to be constant, we find down to $\mb \approx
   16\fm0$ the observed disk fraction to decline much stronger per
   one-magnitude-interval than what would be expected from artificial dimming
   (see Appendix \ref{sec:dimfraction}).

   However, still a fair part of the
   decline \emph{is} likely to be caused by the latter effect: the curve for objects
   dimmed by 1 magnitude lies at about a factor of 1.2 lower than the
   original one, and the two-magnitude curve is even a factor of two
   lower. This shows that our above estimate of the true
   number of dEdis being larger by 1.5 than what we observe is a
   useful estimate for the average fraction of missed objects.

   Still, the issue might be more subtle: if the \emph{relative
   strength} of the disk features was decreasing
   with magnitude in addition to the SNR of the object \emph{as a whole},
   the estimate from artificially dimming the galaxies would be
   somewhat too high. While several of
   the disk features of the artificially dimmed galaxies 
   would still be strong enough to be seen, some of the true observed ones
   would not.
  We examine this possibility in Appendix \ref{sec:dimflat}, and find
   that indeed somewhat more dEdis than
   estimated above might be missed
   at fainter magnitudes due to data limitations. However, if the true
   disk fraction were to decrease to 
   zero this effect would be of minor relevance.
   Although we are not able to give an accurate 
   estimate of the true number fraction of dEdis, we
   point out again that our analysis is consistent with
   the approximation of multiplying the disk fraction with 1.5 in the lower
   panel of Fig.\ \ref{fig:maghist}.
A significantly larger factor can
   be excluded following the argument given in Sect.\
   \ref{sec:sub_flatgal}: the flattening distribution of the brightest
   one-(two-)magnitude-interval of our sample is inconsistent with all
   bright dEs being disk galaxies and instead requires a significant
   number of objects to be spheroids.

   
   \section{Spatial distribution}
 \label{sec:spatial}

\begin{figure}
  \epsscale{0.45}
  \plotone{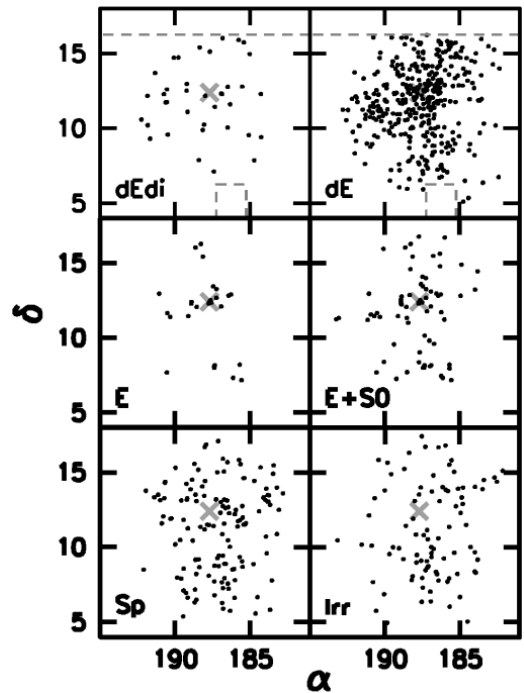}
  \caption{{\bf Distribution of morphological types within the
      cluster.}
    For
    various types of galaxy (dEdis, dEs with no disk features, Es, Es+S0s,
    spirals, and irregulars) the projected spatial distribution is
    shown. Coordinates are given for J2000. Only certain cluster members
    are considered. The position of M87 is shown as grey cross. In the
    upper panels, boundaries of the SDSS coverage are shown as grey
    dashed lines.
  }
  \label{fig:distrmulti}
\end{figure}

   It is well known that the projected spatial distribution of
   different morphological types of galaxies differs significantly
   \citep[the so-called morphology-density
   relation,][]{dre80}. Therefore it appears 
   interesting to examine the distribution of dEdis and dEs where no disk was
   found and compare it to other galaxy types.
   Those projected spatial distributions are shown in Fig.\
   \ref{fig:distrmulti}, along with the
   distributions for giant ellipticals (Es), for Es and giant S0s together,
   for spiral galaxies, and for irregulars for comparison. Positions
   are taken from the VCC by use of 
   the VizieR database \citep{vizier}. Only certain cluster members are
   considered, and intermediate or uncertain classifications between
   the types are excluded\footnote{For example, a galaxy classified as
   'E/S0' is excluded from the sample of Es, but included in the combined
   sample of Es and S0s.}
   Clearly, dEdis show the least clustering of all types, 
   somewhat similar to the distribution of irregulars with
   $\delta>10\arcdeg$.

\begin{figure}
  \epsscale{0.45}
  \plotone{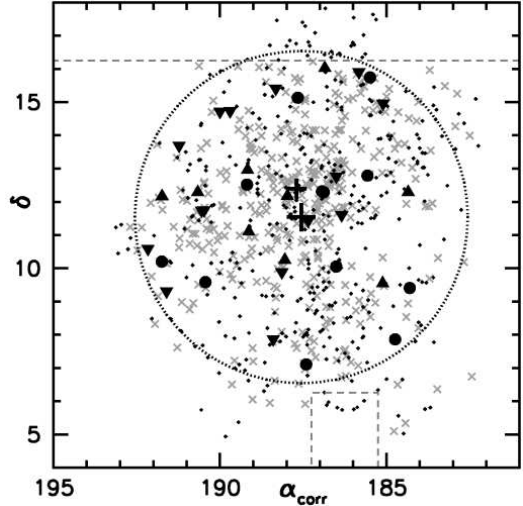}
  \caption{{\bf Distribution of dEdis within the cluster.} 
     Coordinates are given for J2000, and right ascension is corrected
     for the factor $\cos(\delta)$, see text.
     Black circles are certain
     dEdis, black upward-pointing triangles are probable dEdis, and
     black downward-pointing triangles are possible dEdis.
     Grey crosses represent dEs where no disk was found. All other
     Virgo cluster galaxies with $\mb\le18\fm0$ are shown as small
     black dots. Only certain cluster members are considered.
     The upper black cross gives the position of M87, the lower black
     cross marks our cluster center, chosen such that the radius of a
     circle enclosing all dEdis (dotted black line) is minimized
     ($r=5\fdg0$).
     Boundaries of the SDSS coverage are shown
     as grey dashed lines.
  }
  \label{fig:distrnum}
\end{figure}

   For a more quantitative analysis, we show
   the cumulative distribution of each type of galaxy with
   respect to the distance from the cluster center. Since there is no
   unique definition for the latter, we decided to choose a point such
   that the radius of a circle enclosing all dEdis is minimized (Fig.\
   \ref{fig:distrnum}). For this purpose we
   use a 'corrected right ascension', which we define as
   \begin{equation}
     \alpha_{corr} = (\alpha-\alpha_{center})\cdot\cos(\delta) +
   \alpha_{center}\quad ,
   \end{equation}
   so that $\alpha_{corr}$ is measured in true
   degrees. We choose 'our' center to lie at
   \begin{equation}
     \alpha_{center} = \alpha_{M87} - 0\fdg15,\quad \delta_{center} = \delta_{M87} -
   0\fdg85
   \end{equation}
   i.e.\ going from \object{M87} slightly towards M86 and
   M49. Interestingly, this circle at the same time encloses exactly
   all the giant ellipticals. For all other types, we
   only consider galaxies up to the maximum radius of the dEdis, in
   order to properly compare their clustering properties \emph{within}
   that area. The fact that other galaxy types extend slightly further
   outwards might have physical significance, but could also be just due to
   the relatively small number of dEdis as compared to other
   types. Also, a part of it is due to the boundaries of the SDSS DR4
   coverage, indicated with dashed grey lines in the upper panels of
   Fig.\ \ref{fig:distrmulti}.

\begin{figure}
  \epsscale{0.375}
  \plotone{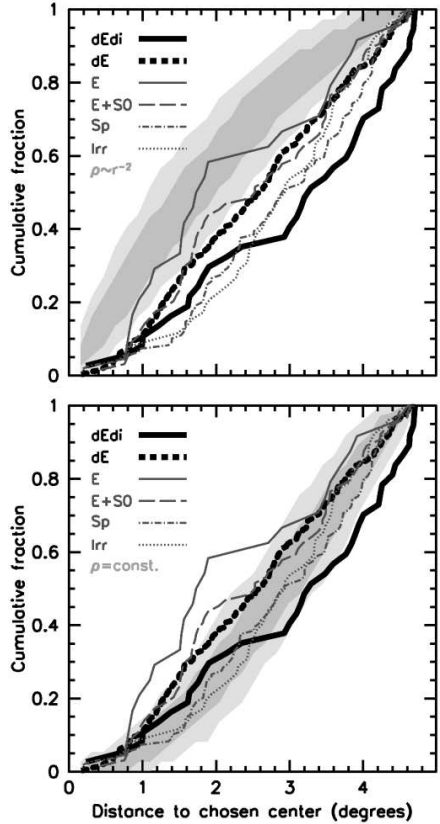}
  \caption{{\bf Radial distribution of morphological types.}
    Both panels show the cumulative distribution of the angular
    distances of galaxies from our chosen cluster center (see
    text). Only certain cluster members are considered, and all
    galaxy types are only considered up to the maximum distance of the
    dEdis. Various line types give the cumulative distributions for
    dEdis, dEs with no disk features, Es, Es+S0s, spirals, and
    irregulars, as labelled in the figure.
    In the upper panel, Monte-Carlo-simulations were performed to yield
    the expected distribution for an isothermal sphere potential
    (i.e.\ $\rho\sim r^{-2}$) for a total number of 36 objects, i.e.\
    the number of (candidate) dEdis. For the simulation, a distance to
    the Virgo cluster center of $15.85\,\rm{Mpc}$ was adopted
    (corresponding to $m-M=31$), resulting in an angular scale of
    $0.28\,\rm{Mpc/\arcdeg}$. The simulated objects populate a sphere
    with a physical radius of $1.4\,{\rm Mpc}$, i.e.\ corresponding to
    the angular value of $5\fdg0$ for the circle in Fig.\
    \ref{fig:distrnum}.
    The resulting distribution is shown as grey
    areas that enclose vertical intervals around the median,
    containing all but
    $\pm 15.87\%$ of simulated values ($\equiv 1\,\sigma$, darker grey)
    and all but $\pm 2.27\%$ ($\equiv 2\,\sigma$, lighter grey).
    In the lower panel, analogous Monte-Carlo-simulations were done for a
    constant galaxy density. 
Note that 1(2)-sigma areas are only valid for a comparison with the
    dEdis, not with other types, since the number of galaxies is
    different for the latter.
  }
  \label{fig:cumnum}
\end{figure}

   We show the cumulative distributions in Fig.\
   \ref{fig:cumnum}. Along with the distribution for different
   morphological types, we show the expected distribution for an
   isothermal sphere ($\rho(r)\sim r^{-2}$) in the upper panel, and for
   constant density ($\rho(r)=\rm{const.}$) in the lower panel, where
   $\rho(r)$ denotes the true volume density, not the projected surface
   density. This is
   done by populating a (three-dimensional) sphere at the distance of
   the Virgo cluster (taken to be $d=15.85\,\rm{Mpc}$, i.e.\
   $m-M=31\fm0$) with the same number of objects as the number of
   dEdis, and then 'observing' the projected distribution of this
   sphere. Vertical intervals containing all but $\pm 15.87\%$ ($\equiv
   1\,\sigma$) of Monte-Carlo-simulated values (darker grey) and all but
   $\pm 2.27\%$ ($\equiv 2\,\sigma$, lighter grey) are shown. Although a sphere is
   clearly not an ideal representation of the dynamically young and
   unrelaxed Virgo cluster, this
   simple model is intended to give at least a rough idea of the actual
   density distribution of the various galaxy classes.
   
   The well-known
   difference in the distribution of giant ellipticals and spirals or
   irregulars is clearly visible, and
 serves as guidance for the
   question of what constitutes a \emph{significant} difference between
   two galaxy types in the diagram.
 The dEs where no disks were found
   roughly follow the distribution of giant Es and S0s, i.e.\ they are
   less centrally clustered than the Es alone but more strongly than spirals
   and irregulars. In contrast, dEdis lie clearly below this
   distribution, and for most of the sample they show even less
   clustering than spirals and irregulars, confirming the impression
   given by Fig.\ \ref{fig:distrmulti}. While giant Es tend towards the
   isothermal sphere and spirals and irregulars
   more or less follow the distribution for constant volume density,
   dEdis even lie beyond the latter -- a clear sign for them being not
   yet virialized, and thus being a population that has experienced
   fairly recent cluster infall.

\begin{figure}
  \epsscale{0.375}
  \plotone{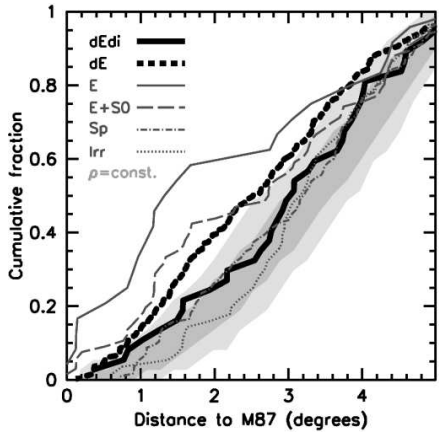}
  \caption{{\bf Radial distribution with respect to M87.}
    Same as in Fig.\ \ref{fig:cumnum}, but now adopting M87 as cluster
    center.
  }
  \label{fig:cumnum_m87}
\end{figure}

   For the sake of completeness, we also
   show the resulting distributions when M87 is chosen as cluster
   center instead (Fig.\ \ref{fig:cumnum_m87}). The difference between
   dEdis and dEs is now slightly less pronounced, but also it now varies
   somewhat less with radius than before. Here the dEdis follow closely
   the distribution of spirals, and fall within the 1-sigma area of
   the theoretical distribution for constant volume density -- note,
   however, that a sphere around M87 is clearly no good representation
   of the Virgo cluster's shape. In contrast to the dEdis,
   dEs where no disk features were found approach the distribution
   of E+S0s, and at larger distances reach the distribution of the Es
   alone.

   
   \section{Discussion and summary}
\label{sec:discuss}

   It is a long-standing question how early-type dwarf galaxies form, and
   whether there is more than one formation channel producing them. Current
   theories include ram-pressure stripping, galaxy harassment, or
   in-situ formation. However, for a proper theoretical approach of dE
   formation, first the characteristics and possible subpopulations of
   the dE class need to be fully understood and unveiled from the
   observational side. While the definition of the dS0 class by
   \citet{vcc} implied a disk nature of these objects,
 the fairly
   diverse classification criteria had to remain suggestive but not
   compelling for dS0s being disk galaxies.
   The discovery of disk features in a handful of dEs
   had not yet been succeeded by a 
   systematic, quantitative study, and could thus not provide significant
   input for models of dE formation.

\begin{figure}
  \epsscale{0.45}
  \plotone{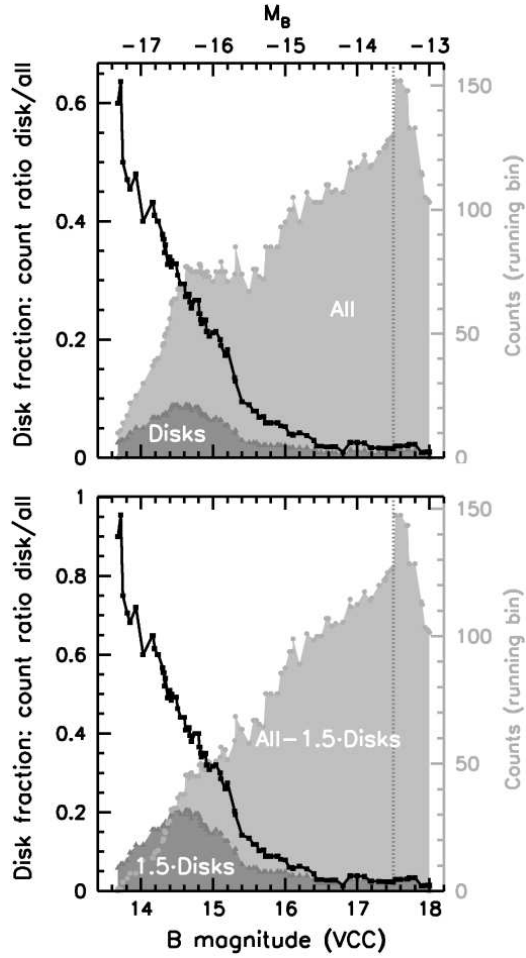}
  \caption{{\bf Luminosity function and disk fraction.}
    {\it Upper panel:} Running histogram of the number of 
    all dEs (light grey) and (candidate) dEdis (dark grey) with respect to
    B-band magnitude as given by the VCC. The bin-width is $1\fm0$, therefore
    the counts are incomplete for $\mb>17\fm5$ (vertical dotted line). A bin is calculated at
    each position of a galaxy in the full sample. The upper x-axis gives
    absolute magnitude assuming $m-M=31\fm0$. Only certain cluster members are
    considered. The ratio of both histograms is the disk fraction and is given
    as black symbols.
    {\it Lower panel:}
    Similar to the upper panel, but for all dEs minus 1.5 times the number of
    (candidate) dEdis (light grey), for 1.5 times the number of (candidate)
    dEdis (dark grey), and for the disk fraction resulting therefrom
    (black).
  }
  \label{fig:maghist}
\end{figure}

\begin{figure}
  \epsscale{0.45}
  \plotone{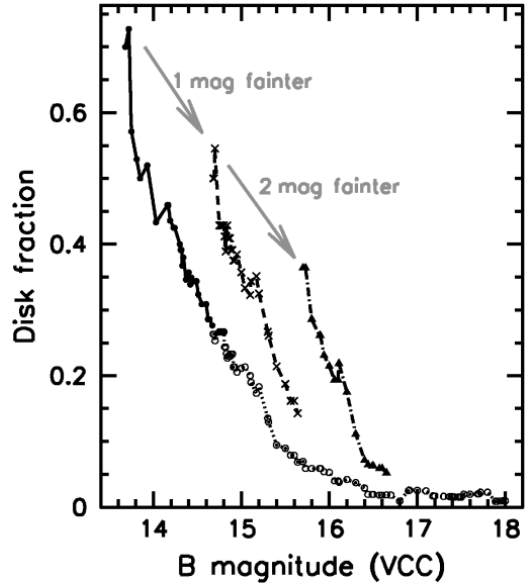}
  \caption{{\bf Effect of SNR on the disk fraction.}
    {\it Upper panel:} Running histogram of the disk fraction as given
    in the upper panel 
    of Fig.\ \ref{fig:maghist} (solid and dashed line with
    circles). The brightest one-magnitude-interval is shown as solid line
    with filled circles, changing to a dotted line with open circles
    outside of the interval. When the dEdis in this interval are dimmed
    (see text) by 1 magnitude, the resulting disk fraction is given by
    the dashed line with crosses. A dimming by 2 magnitudes results in
    the dotted-dashed line with triangles. A histogram bin is calculated at
    each position of a galaxy in the full sample.
  }
  \label{fig:magfaint1}
\end{figure}

 Moreover, kinematics -- which might
   provide further insight into the presence of disks --
 are well
   studied only for a relatively small sample of dEs.
   With the
   SDSS data at hand, we performed for the first time a systematic
   search for disk features in an almost-complete 
   sample of dEs down to $\mb\le 18\fm0$, and found 41 out of 476 objects to
   show (possible, probable) disk features. In the light of the
   diversity of the early-type dwarfs, one of
   our primary and most important results is that dEdis most likely constitute
   a different galaxy population than dEs where no disk features are found:
   the bump in the luminosity function of dEs (Fig.\
   \ref{fig:maghist}) is highly unlikely to be an intrinsic
   characteristic of just a single population, and it is nicely explained
   by the superposition of dEdis and dEs with no disk
   features. Therefore, at least 
   two different formation scenarios appear to be required: one for
   each dEs with and without disk features.

\begin{figure}
  \epsscale{0.45}
  \plotone{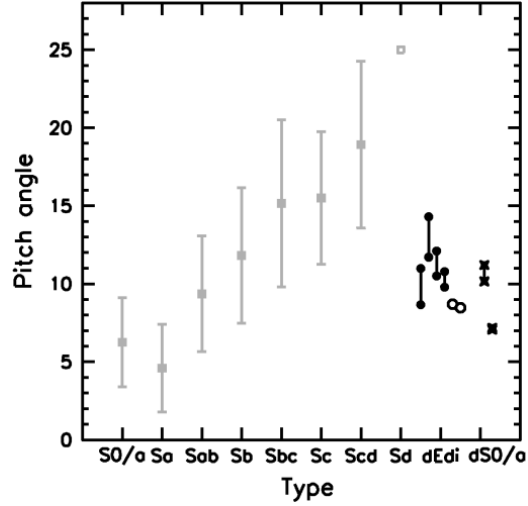}
  \caption{{\bf Pitch angle vs.\ morphology.}
    {\it Grey symbols:} Mean pitch
    angle and 1-$\sigma$ error bars of the spiral arms for various Hubble
    types as given by \citet{ma99}. The value for type Sd was only derived
    from two objects.
    {\it Black filled circles:} Pitch angle for the certain
    cluster members VCC0308, VCC0490, VCC0856, VCC1896 (left to right) for two
    spiral arms each (connected symbol pairs).
    {\it Black open circles:}
    Pitch angle for the possible cluster members VCC0278 (left) and
    VCC1671 for one spiral arm each; in both cases the other arm could
    not be traced well enough.
    {\it Black asterisks:} Pitch 
    angle for the dwarf-like S0/Sa galaxies (see text for details)
    VCC0522 and VCC1902 for two spiral arms each (connected symbol
    pairs; the values for the arms
    of VCC1902 are almost equal).
  }
  \label{fig:pitch}
\end{figure}

   When the first observations of spiral
   structure in dEs were made, galaxy harassment seemed to provide a simple
   explanation for the apparently embedded disks in dwarf ellipticals:
   \citet{mas05} showed
   that the progenitor galaxy's disk need not be completely destroyed during
   the process of transformation, but part of it is left over inside the newly
   formed dwarf. However, we point out a main problem with this scenario: how
   could the observed \emph{well-defined, early-type} spiral-arm structure of dEdis be
   reconciled with them having \emph{late-type} progenitor spirals with
   their typically flocculent arm structure? Figure
   \ref{fig:pitch} directly compares the pitch angle of our objects
   to that of Scd and Sd galaxies in the diagram - and shows an obvious
   mismatch. If one assumes 
   a relatively weak spiral structure for the late-type progenitor that
   would quickly disappear after star formation ceases, one might conclude
   that the above harassment scenario could still be valid, provided that
   the dEdi spiral structure is purely of tidal origin, as e.g.\
   suggested by \citet{jer00a}. The above
   question then changes into asking whether such well-defined spiral arms 
   can at all be created through a process like harassment, and what
   parameters determine their appearance.
 To confirm that we are not looking at spiral structure traced by
 regions of star formation, we examined near-infrared H-band images for
 VCC0308 and VCC0856 which we obtained through the ESO/ST-ECF Science Archive
 facility\footnote{Observations made with ESO/NTT at the La
 Silla Observatory under programme ID 64.N-0288.}. These images show
   the very
 same spiral structure as the optical data, consistent with what would
   be expected for grand-design, early-type spiral arms. A detailed
   examination of the colour properties of the spiral structure will be
   presented in paper III of this series (Lisker et al., in prep.).

\begin{figure}
  \epsscale{0.6}
  \plotone{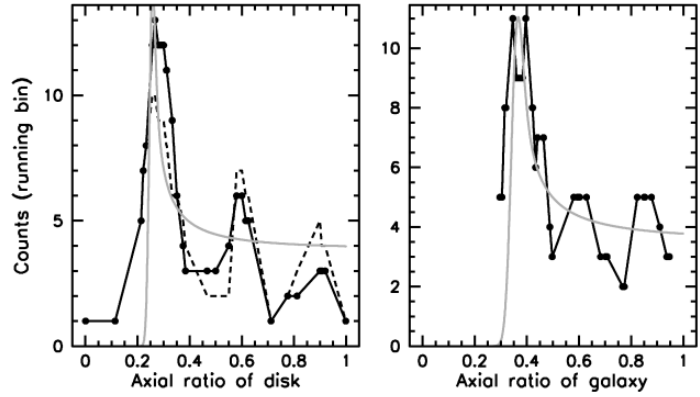}
  \caption{{\bf Axial ratio distribution for disks and galaxies.}
    Running histogram with bin-width 0.1 for all 36 dEdis with certain
    cluster membership.
    {\it Left panel:} Distribution of axial ratio
    measurements of disk features as illustrated in Fig.\
    \ref{fig:incl}. For the solid black line we assume that all elongated
    features where we could not decide between an inclined disk or a bar
    actually are an inclined disk. For the dashed black line, we assume that
    these features are bars, and thus adopt the axial ratio of the
    \emph{galaxy} as an upper limit. The grey line shows the theoretical
    distribution for an intrinsic axial ratio represented by
    a narrow Gaussian of $\mu=0.25$, $\sigma=0.01$, following
    \citet{mihalasbinney}. It is normalized to the same area under the
    curve as the black solid line.
    {\it Right panel:} Distribution of axial
    ratios of the galaxies. The grey line represents an
    intrinsic axial ratio that follows a
    a narrow Gaussian of $\mu=0.35$, $\sigma=0.02$, and is normalized
    like above.
  }
  \label{fig:axgaldisk}
\end{figure}

   Even without considering a specific formation theory, our data also allow us
   to address the question of whether dEdis are genuine disk galaxies
   or whether they are
   spheroids hosting a disk. Our distribution of axial ratios for the
   \emph{disks} (i.e.\ where we measured the disk features, not the
   galaxies as a whole)
   agrees well with the expected distribution assuming an intrinsic
   axial ratio of 0.25 (left panel of Fig.\ \ref{fig:axgaldisk}),
   confirming our general approach to finding disk features in
   dEs. More importantly, also the distribution of axial ratios of the 
   \emph{galaxies} where disk features were found is nicely consistent with
   the assumption of their being disk galaxies with an intrinsic axial
   ratio of 0.35 (right panel of Fig.\ \ref{fig:axgaldisk}). This distribution
   significantly differs from the distribution of dEs 
   with no disk features, the latter being consistent with a distribution of
   genuine spheroids. We do not see how this could be reconciled with the
   assumption that dEdis themselves are spheroids -- instead, we take
   these results as compelling indication for dEdis being disk
   galaxies, represented by the edge-on view of \object[VCC]{VCC1304} (third row of Fig.\
   \ref{fig:pics}). 

   Could this population of disk galaxies be simply an extension of
   their giant counterparts? The deduced intrinsic thickness of dEdis
   (0.35) agrees with the corresponding value for giant Sa
   galaxies as given by \citet{fou90} \citep[0.37 for 'S0/Sa', 0.33 for
   'Sa';][]{sch95}, and the measured pitch angles best agree with
   Hubble type Sab/Sb (Fig.\ \ref{fig:pitch}).
   The dwarf-like S0/Sa galaxies
   presented in Sect.\ \ref{sec:sub_Sa} could in fact bridge the gap
   from dEdis to giant disk galaxies. Here, the presence or absence
   of a 'classical' bulge can distinguish between what would be called
   a giant or a dwarf galaxy.
   However, early-type dwarfs are rare in the field environment, while
   early-type spiral galaxies are preferentially found in the
   field. This fundamental observation provides evidence against
   a close relation of dEdis and early-type spirals.

   The projected
   spatial distribution of dEdis within the Virgo cluster differs
   significantly from dEs with no disk features, and implies that the
   population of dEdis is not virialized yet. Thus, if dEdis would be
   the result of a morphological transformation, this should have occured
   recently. Any spiral structure of the late-type progenitor galaxies
   would have had to be destroyed during the process, since the spiral
   arm characteristics of the dEdis are incompatible with being
   remainders from Sc/Sd spiral galaxies. While a pure star formation
   origin of the spiral arms is unlikely (see above), they might
   originate from the recent galaxy-galaxy interaction that triggered the
   transformation process. Since such spiral structure would quickly
   disappear after the interaction ended, one would expect the dEdis
   to still show structural distortions, i.e.\ to be less homogeneous in
   appearance. Moreover, a significant amount of tidal debris should
   still be present around them. At least the latter issue could be settled
   observationally with dedicated deep imaging of dEdis and their vicinity.
   
\begin{figure}
  \epsscale{0.45}
  \plotone{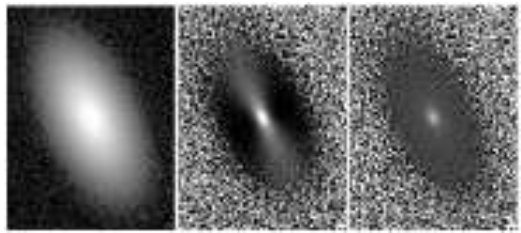}
  \caption{{\bf Circular and elliptical unsharp masks.}
    {\it Left panel:}
    Simulated galaxy image created with \emph{IRAF\,/\,mkobjects}, with 
    exponential intensity profile, scale length along major axis 20 pixels, and
    axial ratio 0.5. {\it Middle Panel:} 'Circular' unsharp mask of the
    simulated galaxy, created with
    a circular Gaussian of kernel size $\sigma=4\,\rm{pix}$.
An elongated feature appears due to the application of a
    circular Gaussian to an elliptical object.
    {\it Right Panel:} 'Elliptical' unsharp mask with the same kernel size along the
    major axis, created with an elliptical Gaussian
    matching the position angle and axial ratio of the galaxy.
      Each panel has a horizontal scale of $138$ pixels.
  }
  \label{fig:simul}
\end{figure}

\begin{figure}
  \epsscale{0.4}
  \plotone{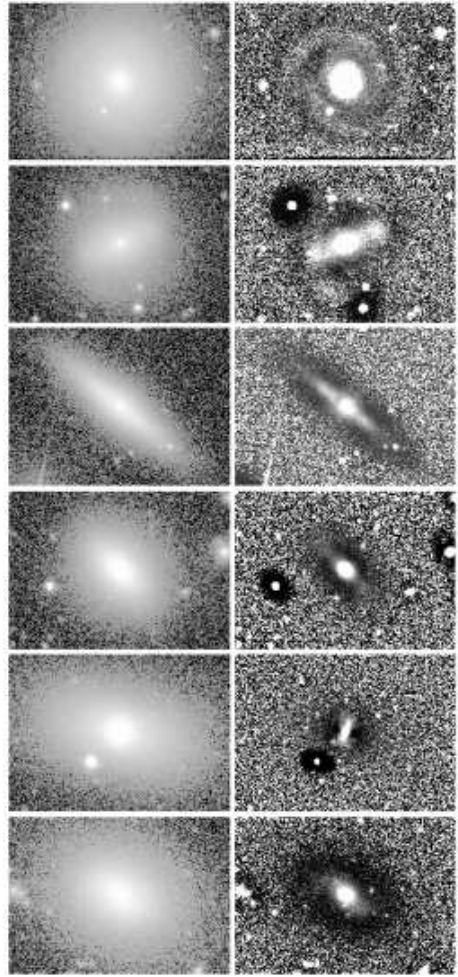}
  \caption{{\bf Early-type dwarfs with disk features: dEdis.}
    Combined
  images and unsharp masks for three dEs with unambiguous disk features
  (top three rows), one probable dEdi (fourth row), and two possible
  dEdis (last two rows). The galaxies are, from top to bottom: VCC0308
  (spiral arms; unsharp mask kernel size $\sigma=20\,\rm{pix}$), VCC1896
  (bar and weak spiral arms; $\sigma=13\,\rm{pix}$), VCC1304 (edge-on
  disk; $\sigma=20\,\rm{pix}$), VCC0990 (inclined disk, also see Fig.\
  \ref{fig:simul}; $\sigma=9\,\rm{pix}$), VCC1183 (bar;
  $\sigma=6\,\rm{pix}$), VCC2019 (possibly inclined disk, maybe warped or
  distorted; $\sigma=13\,\rm{pix}$).
      Each panel has a horizontal scale of $98''$ ($7.55\,$kpc  with
    $d=15.85\,\rm{Mpc}$).
  }
  \label{fig:pics}
\end{figure}

   Even before the discovery of the first spiral structure within a dE
   by \citet{jer00a}, it was obvious from the existence of a dS0
   class that treating all early-type dwarfs
   as one single population of galaxies always bore the risk of mixing
   objects that might have had different evolutionary
   histories. With our systematic search
   for disk features, we have now provided several strong indications that
   early-type dwarfs do indeed consist of two distinct populations of
   galaxies. Therefore, with our results at hand, we strongly recommend
   that those objects identified by us as (candidate) dEdis be
   considered separately from the rest of dEs in any future study of
   early-type dwarfs, like e.g.\ a study of dE colours. Furthermore,
   one should keep in mind that a significant fraction of the brighter
   dEs where we did not \emph{find} any disk features might still \emph{be}
   dEdis -- this possible incompleteness could fake systematic differences between
   brighter and fainter dEs. We also suggest
   to separately consider objects where we did not find
   disk features but that have been classified as dS0 in the VCC, since
   our results confirm that these also differ from 'ordinary' dwarf
   ellipticals.
    As a technical recommendation, we advise caution on the
    interpretation of substructure that is seen in unsharp mask images
    created with \emph{isotropic} smoothing of a \emph{non-circular}
   object: as illustrated in Fig.~\ref{fig:simul}, this can lead to
   artificial elongated features similar to an edge-on disk.

   Now that the separation between dEs and dEdis has been established,
   their properties can be analyzed. Given the disk nature of the
   dEdis, a correlation with kinematical studies of early-type dwarfs
   is an obvious thing to do.
   Such a correlation has first been investigated by \citet{geh03}
   who found that two out of three rotating dEs show disk features, yet 
   two out of four non-rotating dEs have weak disk substructure as
   well.
 Without going into the details of the
   kinematical analyses, we compiled results from several studies that
   state whether or not a dE shows significant rotation
   \citep{vZe04a,geh03,sim02}. Note that these studies differ in their
   data properties, their maximum radius for sampling the rotation
   curve, and their criteria for significant rotation.
18 out of 29
   galaxies are found to be not rotationally supported, i.e.\ they show
   no or too slow rotation as compared to the observed velocity
   dispersion. 4 of these objects (22\%) are (candidate)
   dEdis. However, it needs to be stressed that rotation curves are
   only sampled out to about the half-light radius, which might not be
   enough for definite statements about rotational support. 3 of those 4
   dEdis do show significant rotation, but not enough to qualify for being
   rotationally supported.
   Of the
   11 galaxies that were found to be rotationally supported, 6 (55\%)
   are (candidate) dEdis. There is thus a
   tendency for dEdis to be rotationally supported systems, as one would
   expect for disk galaxies. The number statistics are consistent 
   with our rough estimate of a third of the dEdis being missed in our
   study when assuming that most or all of them are rotationally supported.

   Given the different spatial distribution of dEdis and dEs within the
   cluster, a further issue of interest would of course be
   their distribution of heliocentric velocities. These are available
   for 31 dEdis and 162 dEs where no disks were found. However, the two
   distributions do not differ significantly. Since the true
   threedimensional locations of our galaxies within the cluster are
   not known, let alone the exact threedimensional structure of the
   cluster itself, unfortunately no useful conclusion can be drawn
   here.

\begin{figure}
  \epsscale{0.45}
  \plotone{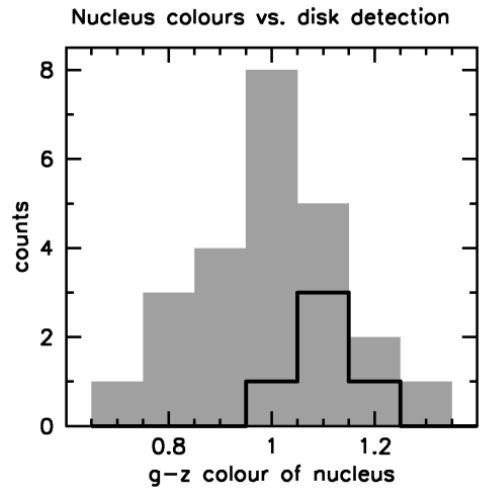}
  \caption{{\bf Nucleus colours of dEdis.}
    Histogram of g-z colours of dE nuclei (grey) as derived by
    \citet{str05}. Five of these objects are dEdis; their nucleus
    colours are shown as black histogram.
  }
  \label{fig:strader}
\end{figure}

   To demonstrate how our recommended separation of dEdis and the rest
   can be applied to other studies of early-type dwarfs, we show in Fig.\
   \ref{fig:strader} the colours of dE nuclei derived by \citet{str05}:
   five objects of this sample are identified by us as dEdis, and
   show redder nucleus colours than the bulk of dEs. To obtain a
   clearer relation, it would be desirable to further pin down the
   possible disk nature of the remaining dEs where we could not find
   disk features. This calls for a larger sample of kinematically studied dEs
   as well as for deeper images of higher resolution to detect further
   substructure, so that more quantitative input for theories
   of dE and dEdi formation can eventually be provided.


\acknowledgements
    We gratefully acknowledge support by the Swiss National Science
    Foundation through grant number 200020-105260. We thank the referee
    for constructive suggestions.
    T.L.\ would like to thank Victor Debattista for repeatedly
    affirming that dwarf galaxies \emph{are} interesting.
    This study would not have been possible without the wealth of publicly
    available data from the SDSS Data Release 4.    
    Funding for the SDSS has been provided
    by the Alfred P. Sloan Foundation, the Participating Institutions,
    the National Aeronautics and Space Administration, the National
    Science Foundation, the U.S. Department of Energy, the Japanese
    Monbukagakusho, and the Max Planck Society. The SDSS Web site is
    {\it http://www.sdss.org/}. 
    The SDSS is managed by the Astrophysical Research Consortium (ARC)
    for the Participating Institutions. The Participating Institutions
    are The University of Chicago, Fermilab, the Institute for Advanced
    Study, the Japan Participation Group, The Johns Hopkins University,
    the Korean Scientist Group, Los Alamos National Laboratory, the
    Max-Planck-Institute for Astronomy (MPIA), the Max-Planck-Institute
    for Astrophysics (MPA), New Mexico State University, University of
    Pittsburgh, University of Portsmouth, Princeton University, the
    United States Naval Observatory, and the University of Washington. 
    This research has made use of NASA's Astrophysics Data
    System Bibliographic Services, and of the NASA/IPAC Extragalactic
    Database (NED) which is operated by the Jet Propulsion Laboratory,
    California Institute of Technology, under contract with the
    National Aeronautics and Space Administration.



\mbox{~}

\clearpage

\begin{deluxetable}{llllll}
  \tablecaption{Early-type dwarfs with disk features. \label{tab:disks}}
  \tablehead{
    \colhead{VCC} & \colhead{$m_{\rm B}$}
    & \colhead{$\alpha_{\rm J2000}$} & \colhead{$\delta_{\rm J2000}$}
    & \colhead{Membership} & \colhead{Note}
  }
  \startdata
\sidehead{Certain disks}
 1010 & 13\fm72 & 12$^{\rm h}$27$^{\rm m}$27\fs4 & +12\arcdeg17\arcmin25\arcsec & M &  3,4,(5)\\
 0523 & 13.75 & 12 22 04.1 & +12 47 15 & M &  3,4,(5)\\
 2048 & 13.85 & 12 47 15.3 & +10 12 13 & M &  1      \\
 1036 & 14.03 & 12 27 41.2 & +12 18 57 & M &  2      \\
 0308 & 14.30 & 12 18 50.9 & +07 51 43 & M &  5      \\
 0490 & 14.33 & 12 21 38.8 & +15 44 42 & M &  5      \\
 0856 & 14.42 & 12 25 57.9 & +10 03 14 & M &  5      \\
 1695 & 14.60 & 12 36 54.9 & +12 31 12 & M &  1,5    \\
 1896 & 14.78 & 12 41 54.6 & +09 35 05 & M &  3,5    \\
 1671 & 14.80 & 12 36 32.2 & +06 10 11 & P &  5      \\
 0216 & 14.90 & 12 17 01.1 & +09 24 27 & M &  5,(3)  \\
 0278 & 15.10 & 12 18 14.4 & +06 36 14 & P &  5      \\
 1304 & 15.50 & 12 30 39.9 & +15 07 47 & M &  2      \\
 1204 & 16.60 & 12 29 38.0 & +07 06 24 & M &  2      \\
\sidehead{Probable disks}
 1422 & 13.81 & 12 32 14.2 & +10 15 06 & M &  1      \\
 1949 & 14.19 & 12 42 57.8 & +12 17 14 & M &  2,3,(4)\\
 1947 & 14.56 & 12 42 56.4 & +03 40 36 & P &  3,4    \\
 1392 & 14.62 & 12 31 55.9 & +12 10 28 & M &  2      \\
 0407 & 14.64 & 12 20 18.8 & +09 32 44 & M &  2      \\
 0990 & 14.81 & 12 27 16.9 & +16 01 28 & M &  2      \\
 0218 & 14.88 & 12 17 05.4 & +12 17 22 & M &  2,(6)  \\
 2050 & 15.20 & 12 47 20.6 & +12 09 59 & M &  2      \\
 0336 & 16.20 & 12 19 17.6 & +05 52 33 & P &  1      \\
 1691 & 17.30 & 12 36 51.1 & +12 57 31 & M &  6,(5)  \\
\sidehead{Possible disks}
 1910 & 14.17 & 12 42 08.7 & +11 45 15 & M &  1      \\
 1183 & 14.32 & 12 29 22.5 & +11 26 02 & M &  3      \\
 0389 & 14.40 & 12 20 03.3 & +14 57 42 & M &  4      \\
 2019 & 14.55 & 12 45 20.4 & +13 41 34 & M &  4,(5)  \\
 0608 & 14.70 & 12 23 01.7 & +15 54 20 & M &  2      \\
 2042 & 14.79 & 12 46 38.2 & +09 18 27 & M &  4,(5)  \\
 1779 & 14.83 & 12 39 04.7 & +14 43 52 & M &  2      \\
 1684 & 14.87 & 12 36 39.4 & +11 06 07 & M &  2,(7)  \\
 1836 & 14.92 & 12 40 19.6 & +14 42 55 & M &  5      \\
 0397 & 15.00 & 12 20 12.2 & +06 37 24 & P &  2,4,(3)\\
 1514 & 15.10 & 12 33 37.7 & +07 52 17 & M &  2      \\
 1444 & 15.60 & 12 32 35.9 & +09 53 11 & M &  6      \\
 0788 & 15.80 & 12 25 16.8 & +11 36 19 & M &  2      \\
 1921 & 15.90 & 12 42 26.5 & +11 44 25 & M &  2      \\
 2080 & 16.20 & 12 48 58.4 & +10 35 12 & M &  2      \\
 0854 & 17.30 & 12 25 55.7 & +12 46 11 & M &  6      \\
 1505 & 18.00 & 12 33 24.7 & +15 24 28 & M &  6      \\
  \enddata
  \tablecomments{
    Objects are sorted by B-band magnitude $m_{\rm B}$ as given by
 \citet{vcc}. Cluster membership is provided by
 \citet{vcc,virgokin}: M\,=\,certain cluster member, P\,=\,possible 
    member. The last column contains information about the nature of
  the identified features: 1\,=\,bar or edge-on disk, 2\,=\,inclined disk,
  3\,=\,bar, 4\,=\,disk, 5\,=\,spiral arms, 6\,=\,too flat for a
  spheroid, 7\,=\,central gas or dust. The latter is an additional
  feature, but is not counted as disk. Numbers in brackets give
  uncertain features of which only a hint is present.\\
  }
\end{deluxetable}
 
\begin{deluxetable}{l|lll|lll}
  \tablecaption{Relative strength of spirals. \label{tab:strength}}
  \tablehead{
    \colhead{VCC}
    & \colhead{$\frac{f_{\rm res}}{f_{\rm mod}}$}
    & \colhead{$\Delta m$}
    & \colhead{$\frac{f_{\rm res}}{f_{\rm total}}$}
    & \colhead{$\left(\frac{f_{\rm res}}{f_{\rm mod}}\right)_{\rm smoo}$}
    & \colhead{$\Delta m_{\rm smoo}$}
    & \colhead{$\left(\frac{f_{\rm res}}{f_{\rm total}}\right)_{\rm smoo}$}
}
  \startdata
  0308 & 0.107 & 2.43 & 0.097 & 0.082 & 2.71 & 0.076\\
  0490 & 0.132 & 2.20 & 0.117 & 0.122 & 2.29 & 0.108\\
  0856 & 0.075 & 2.81 & 0.070 & 0.059 & 3.07 & 0.056\\
  \noalign{\smallskip}
  \colrule
  \noalign{\smallskip}
  0522 & 0.159 & 2.00 & 0.137 & 0.127 & 2.24 & 0.113\\
  1902 & 0.150 & 2.06 & 0.131 & 0.102 & 2.47 & 0.093\\
  \enddata
  \tablecomments{~Columns 2-4 give measured values for the optimized
  residual image \emph{without} median smoothing, columns 5-7 give the
  same quantities for the version \emph{with} smooting (see text for
  details). Columns 2 and 5 give the ratio of the flux of the residual
  image to the flux of the model image within the chosen
  aperture. Columns 3 and 6 give the same as a magnitude difference,
  and columns 4 and 7 give the fraction of residual to total
  light.}
\end{deluxetable}

\clearpage

\clearpage
   
\begin{appendix}

   \section{Residual image optimization}
 \label{sec:resiopti}

\begin{figure}
  \epsscale{0.5}
  \plotone{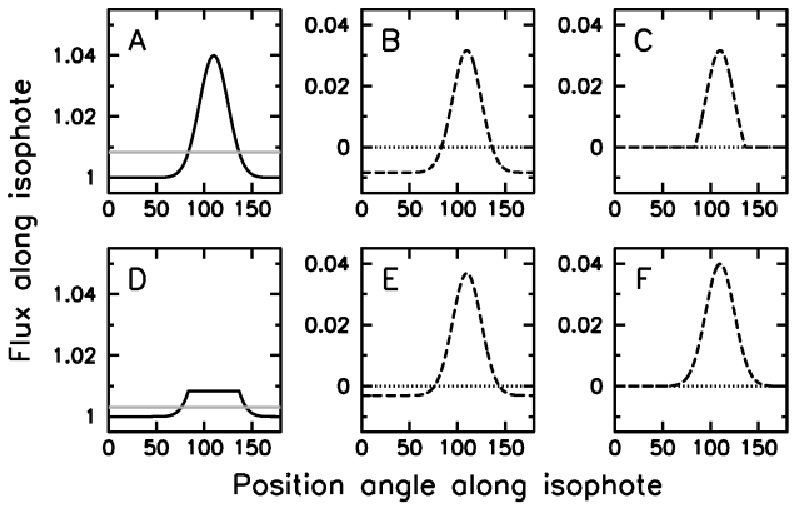}
  \caption{{\bf Residual image optimization.}
    Sketch of the iterative
  method for improving the spiral arm residual image. Each panel shows
  the flux distribution along an elliptical isophote,
  i.e.\ with respect to position angle. The isophote is represented by
    a constant flux value superposed by a crossing spiral arm modeled
    by a Gaussian. See text for the details of the
  method. Panel A shows the initial flux as solid line, with the
    average flux value given as grey line. Panel B shows the residual
    flux as dashed line, with the zero value given as dotted
    line. Panel C results from B when all negative values are set to
    zero. Panel D is obtained by subtracting panel C from A, with the
    new average flux value given as grey line. This value is
    subtracted from the original flux and results in the residual flux
    given in panel E. Panel F shows the final residual
    flux after 9 iterations.
  }
  \label{fig:sketch}
\end{figure}

   In the residual images obtained in Sect.\ \ref{sec:sub_residual},
   the flux level in between the spiral arms is negative:
   when fitting ellipses, the average flux value of 
   each elliptical isophote is affected by the spiral arms and thus comes out
   slightly too high (panel A of Fig.\ \ref{fig:sketch}). This results
   in negative flux values when the model is subtracted from the 
   original image (panel B). We construct optimized residual images through the
   following iterative procedure.
   Where the initial residual image has negative flux values, its flux is set
   to zero, otherwise it is left unchanged (panel C). The resulting image is then
   subtracted from the original galaxy image (panel D), and a new residual image is
   obtained like before by fitting ellipses, constructing a new galaxy
   model, and subtracting it from the original image (panel E). This is
   repeated nine times iteratively, so that the 
   final (tenth) residual image has reached (or come close to) a flux level of
   zero in between the spiral arms (panel F). A slight variation of
   this procedure is to smooth the 
   residual image with a $3\times 3$ pixel median filter each time before the
   negative flux values are set to zero. It turns out that the final image of
   the latter version still has a slightly negative overall flux level, while
   the version without smoothing yields a slightly positive (i.e.\ too high)
   overall value in the residual image. We therefore use the strength
   measurement from the version with smoothing as lower limit, and the one
   without smoothing as upper limit. 


   \section{Artificial dimming of the galaxies}
 \label{sec:dim}

   In order to artificially dim our objects by 1 (2) magnitudes, first
   the object size was
   decreased by a factor of 1.2 per magnitude with \emph{IRAF\,/\,MAGNIFY},
   preserving the total flux. This follows the relation of
   magnitude and radius of the dEs \citep{bin91}: on average, the
   radius decreases with a roughly a factor of 1.2 per magnitude. Since
   this demagnification also affects the PSF, the image 
   was then convolved with a (normalized) Moffat kernel of proper size
   so as to approximately reproduce the original SDSS PSF \citep[taken to be 
   $1\,\rm{FWHM}=4\,\rm{pix}$;][]{sdssedr}. We then added noise to the
   image, with a $\sigma$ larger by 2.51 (6.56) compared to the
   original noise, thereby simulating the SNR of the 1 (2) magnitude fainter
   object. To increase the noise $\sigma$ by 1 magnitude, one
   would actually need to add noise with
   $\rm{\sigma'}= \sqrt{2.51^2-1^2}\cdot
   \rm{\sigma}=2.30\,\rm{\sigma}$. However, since
   the original noise has already been weakened by demagnifying the
   image, we chose to use $\rm{\sigma'}= 2.51\,\rm{\sigma}$ as a
   conservative approximation instead.

   
   \section{Effect of SNR on the disk fraction}
 \label{sec:dimfraction}

\begin{figure}
  \epsscale{0.45}
  \plotone{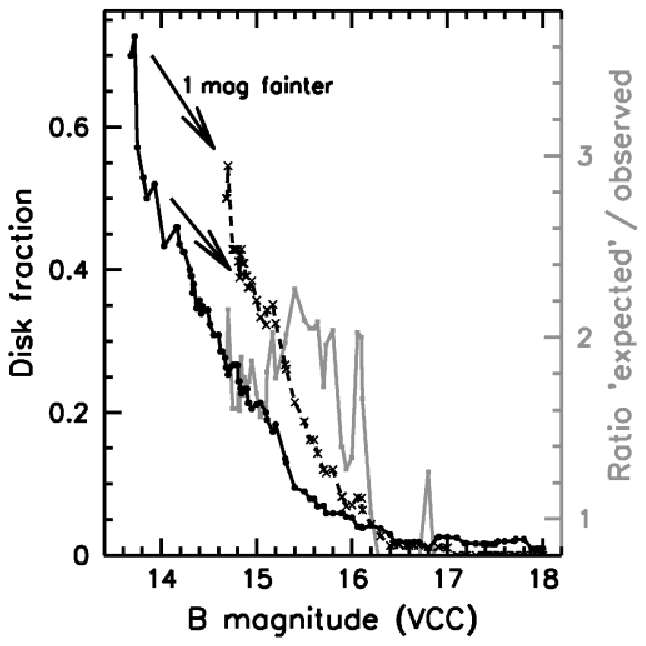}
  \caption{{\bf Decline of the disk fraction.}
      Running histogram of the disk fraction as given
      in the upper panel
      of Fig.\ \ref{fig:maghist} (solid line with circles). When all
      dEdis are dimmed by 1 magnitude (see text), the resulting disk
      fraction is given by the dashed line with crosses. A histogram bin
      is calculated at each position of a galaxy in the full sample. The
      grey line gives the ratio of both running histograms, and
      illustrates how much stronger the observed disk fraction declines
      per one-magnitude-interval than the 'expected' fraction does from
      artificial dimming only.
  }
  \label{fig:magfaint2}
\end{figure}

   In Fig.\ \ref{fig:magfaint2} we show a running histogram of the
   observed disk fraction (solid line with filled circles) and of the
   fraction obtained after dimming all objects by 1 magnitude (dashed
   line with crosses). The original disk fraction lies clearly below the
   shifted one until the region where both become very small and
   are affected by small number statistics. It is important to point
   out that Fig.\ \ref{fig:magfaint2} does \emph{not} show how many dEdis
   would be detected assuming a constant true disk fraction. Instead,
   since all objects are dimmed by an equal amount (namely 1
   magnitude), it shows the disk fraction that we would expect to find at a
   magnitude $m$ when starting from the observed fraction at $m-1$ and
   artificially dimming the objects there. Thus, any difference between the
   observed value at $m-1$ and the 'expected' value at $m$ is due to
   data limitations only. This is symbolized by the arrows in the
   figure. Consequently, if the 
   observed decline from $m-1$ to $m$ is stronger than this 'expected'
   one, at least part of it has to be real and be not only due to data
   limitations. The ratio between the two curves thus tells us how much
   stronger the observed decline \emph{per one-magnitude-interval} is compared
   to what artificial dimming of the galaxies would predict. We plot
   this ratio in Fig.\ 
   \ref{fig:magfaint2} as grey solid line. Until $\mb \approx
   16\fm0$ -- where the number of dEdis becomes very small -- for each
   one-magnitude-step the observed disk fraction 
   declines a factor of 1.5-2.2 stronger than the 'expected' one from
   limitations of our data only. This clearly shows that the decline of
   the disk fraction is real.

   
   \section{Effect of SNR on the flattening distribution of the disks}
\label{sec:dimflat}

\begin{figure}
  \epsscale{0.35}
  \plotone{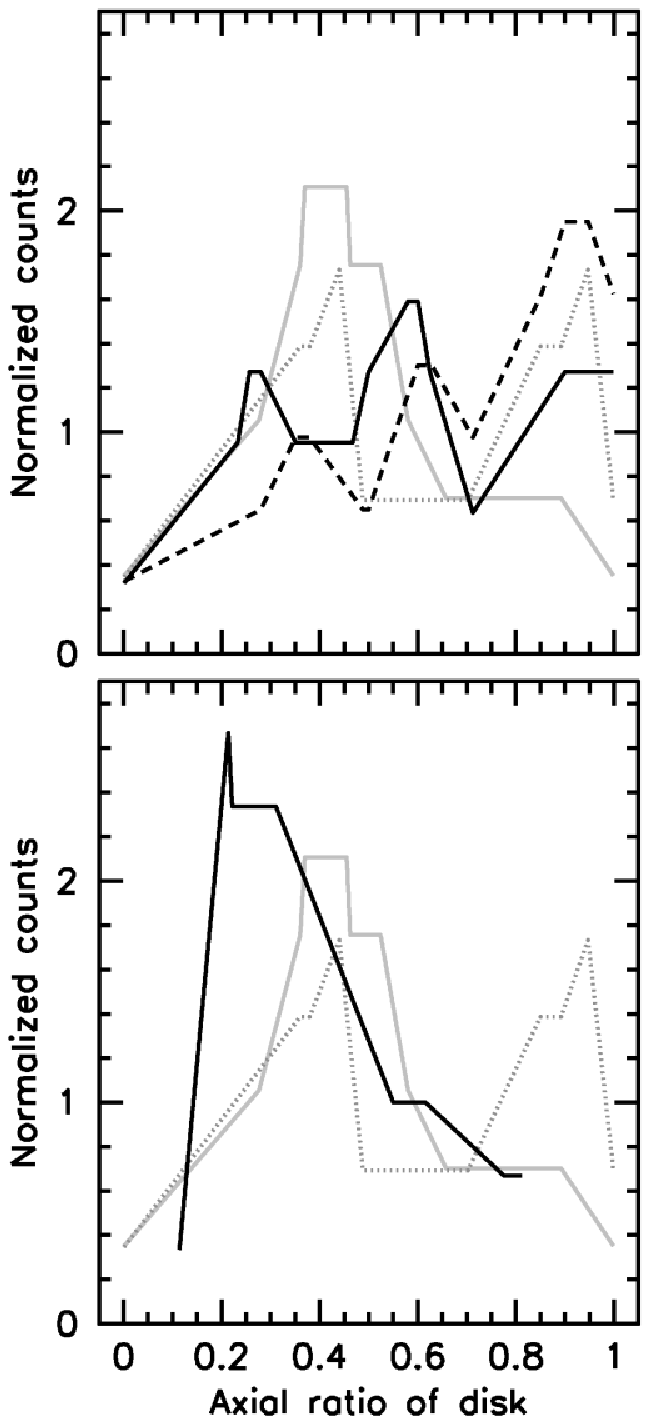}
  \caption{{\bf Effect of SNR on the flattening distribution.}
 {\it Upper panel:} Distribution of axial ratio
  measurements of disk features as illustrated in Fig.\
  \ref{fig:incl}, but this time only for the galaxies in the brightest
 one-magnitude-interval. The black lines show the running histogram of
 the original measurements, while the grey lines give the axial ratios
 measured for the disk features after artificially dimming the galaxies
 by 1 magnitude. A bin is calculated at each data point of each curve with a
 bin-width of 0.2, and the counts are normalized to an area of 1 under
 the curve. {\it Lower panel:} Same as above, but here we compare the
 observed \emph{second-brightest} one-magnitude-interval (black) to the
 \emph{brightest} interval dimmed by 1 magnitude (grey, same as in the
 upper panel).
  }
  \label{fig:axhist}
\end{figure}
 
   Apparent axial ratios of disk features were not only measured on the
   original images, but also on those where the galaxies had been
   artificially dimmed, in order to reveal potential changes in the
   flattening distribution with magnitude.
   In the upper panel of Fig.\ \ref{fig:axhist} we compare the
   distribution of disk axial ratios for both the observed (black) and the
   artificially dimmed (grey) galaxies that lie within the brightest
   one-magnitude-interval of our sample. As in Fig.\ \ref{fig:axgaldisk}, solid
   lines are derived from lower limits of the axial ratios, dashed
   lines from upper limits, depending on the interpretation of an
   elongated feature as a bar or as an edge-on disk. Both
   curves are normalized to an area of 1.
     They show a tendency for the
   dimmed objects towards lower axial ratios, indicating that the
   (artificial) dimming of objects might slightly prefer disks
   of certain inclinations over others. However, the distribution of
   axial ratios for the galaxies of the \emph{observed}
   second-brightest one-magnitude-interval is much more clearly skewed towards
   smaller axial ratios, i.e.\ larger inclinations (solid line in the
   lower panel of Fig.\ \ref{fig:axhist}).
   While the upper panel suggests that part of this is due to the
   effect of the SNR on the detectability of features like spiral arms,
   it might also be that such features are intrinsically weaker -- or
   not even present -- in fainter objects. For example, the observed
   second-brightest one-magnitude-interval does not contain objects
   that look like the close-to-face-on
   spirals in all of \object[VCC]{VCC0308}, \object[VCC]{VCC0490}, and
   \object[VCC]{VCC0856}, although it does contain galaxies with weaker
   spiral features that have a larger inclination. Note, however, that the
   black histogram in the upper panel consists of
     16 objects, and both the grey histogram and the black histogram in
   the lower panel consist of only 13 objects.
   Therefore, the axial ratio distributions
   could at least to some extent be affected by small number
   statistics. We emphasize that the above effects on the
   axial ratio of the \emph{disk} features need not go hand in hand
   with the axial ratios of the \emph{galaxies}: as an example, the
   weak spiral arms in \object[VCC]{VCC1896} are not seen anymore when
   the galaxy is dimmed by one magnitude. One could then confuse
   the bar with being an inclined disk and thus measure a much smaller
   axial ratio of the disk feature, while the galaxy's axial ratio is
   the same in both cases.

   
   \section{Objects where no disk features were found}
 \label{sec:nodisk}
VCC numbers of objects where no substructure was found:\\
0011, 0029, 0033, 0050, 0061, 0065, 0068, 0070, 0082, 0091, 0096, 0106,
0108, 0109, 0115, 0118, 0127, 0158, 0173, 0178, 0200, 0208, 0227, 0230,
0235, 0236, 0244, 0261, 0273, 0287, 0292, 0294, 0299, 0303, 0317, 0319,
0321, 0330, 0335, 0346, 0361, 0372, 0388, 0390, 0394, 0396, 0401, 0403,
0418, 0421, 0436, 0439, 0440, 0444, 0452, 0454, 0458, 0461, 0466, 0499,
0503, 0504, 0510, 0525, 0539, 0542, 0543, 0545, 0554, 0558, 0560, 0561,
0587, 0592, 0594, 0600, 0611, 0622, 0632, 0634, 0652, 0653, 0668, 0674,
0684, 0687, 0695, 0706, 0711, 0723, 0725, 0745, 0746, 0747, 0748, 0750,
0753, 0755, 0756, 0760, 0761, 0762, 0765, 0769, 0775, 0777, 0779, 0786,
0790, 0791, 0795, 0803, 0808, 0810, 0812, 0815, 0816, 0817, 0820, 0823,
0824, 0833, 0838, 0839, 0840, 0846, 0855, 0861, 0862, 0863, 0871, 0872,
0877, 0878, 0882, 0896, 0916, 0917, 0920, 0926, 0928, 0930, 0931, 0933,
0936, 0940, 0949, 0953, 0965, 0972, 0974, 0976, 0977, 0983, 0991, 0992,
0997, 1005, 1028, 1034, 1039, 1040, 1044, 1059, 1064, 1065, 1069, 1073,
1075, 1076, 1079, 1087, 1089, 1092, 1093, 1099, 1101, 1104, 1105, 1107,
1111, 1115, 1119, 1120, 1122, 1123, 1124, 1129, 1132, 1137, 1149, 1151,
1153, 1163, 1164, 1167, 1172, 1173, 1185, 1191, 1198, 1207, 1209, 1210,
1212, 1213, 1218, 1222, 1223, 1225, 1228, 1235, 1238, 1239, 1240, 1246,
1254, 1261, 1264, 1268, 1296, 1298, 1302, 1307, 1308, 1311, 1314, 1317,
1323, 1333, 1337, 1348, 1351, 1352, 1353, 1355, 1366, 1369, 1373, 1384,
1386, 1389, 1396, 1399, 1400, 1402, 1407, 1414, 1417, 1418, 1420, 1430,
1431, 1432, 1438, 1441, 1443, 1446, 1449, 1451, 1453, 1464, 1472, 1481,
1482, 1488, 1489, 1491, 1495, 1496, 1498, 1503, 1509, 1517, 1518, 1519,
1523, 1528, 1531, 1533, 1539, 1549, 1553, 1561, 1563, 1565, 1571, 1573,
1577, 1599, 1601, 1603, 1604, 1606, 1609, 1616, 1622, 1629, 1642, 1643,
1647, 1649, 1650, 1651, 1652, 1657, 1658, 1661, 1663, 1669, 1674, 1677,
1682, 1683, 1687, 1688, 1689, 1702, 1704, 1710, 1711, 1717, 1719, 1729,
1733, 1740, 1745, 1755, 1761, 1762, 1764, 1767, 1773, 1785, 1792, 1794,
1796, 1803, 1806, 1812, 1815, 1826, 1828, 1829, 1831, 1839, 1843, 1857,
1861, 1866, 1867, 1870, 1876, 1879, 1881, 1886, 1887, 1890, 1891, 1893,
1895, 1897, 1901, 1909, 1912, 1915, 1917, 1919, 1928, 1934, 1936, 1942, 1945,
1948, 1950, 1951, 1958, 1964, 1966, 1967, 1971, 1980, 1982, 1983, 1991,
1995, 2004, 2008, 2011, 2012, 2014, 2017, 2028, 2032, 2043, 2049, 2051,
2054, 2056, 2061, 2063, 2074, 2078, 2081, 2083, 2088.

VCC numbers of objects where substructure other than a disk was found
(see Sect.\ \ref{sec:sub_disks}):\\
0009, 0021, 0046, 0170, 0209, 0281, 0288, 0338, 0501, 0636, 0781, 0870,
0929, 0951, 0962, 1078, 1288, 1334, 1370, 1395, 1457, 1501, 1512, 1567,
1617, 1668, 1715, 1743, 2045.


\end{appendix}

\end{document}